\documentclass[11pt,a4paper]{article}

\usepackage{ascmac}
\usepackage{amsmath}
\usepackage{amssymb}
\usepackage{bm}
\usepackage[footnotesize,bf]{caption}
\usepackage{fullpage}
\usepackage{color}
\usepackage{float}
\usepackage{graphicx}
\usepackage[hidelinks]{hyperref} 
\usepackage[]{natbib}
\usepackage{subfigure}
\usepackage{txfonts}
\usepackage{threeparttable}
\usepackage{wrapfig}
\usepackage[]{multicol}
\usepackage{wrapfig}
\usepackage{authblk}

\usepackage{floatflt}

\title{\large Complete list of the ASTRO-H Science Working Group}
\date{\vspace{-0.5cm}}
\newcommand{\MakeWhitePaperTitle}{
	\begin{center}
		\begin{figure}
			\vspace{1cm}
			\begin{center}
				\includegraphics[width=0.2\hsize]{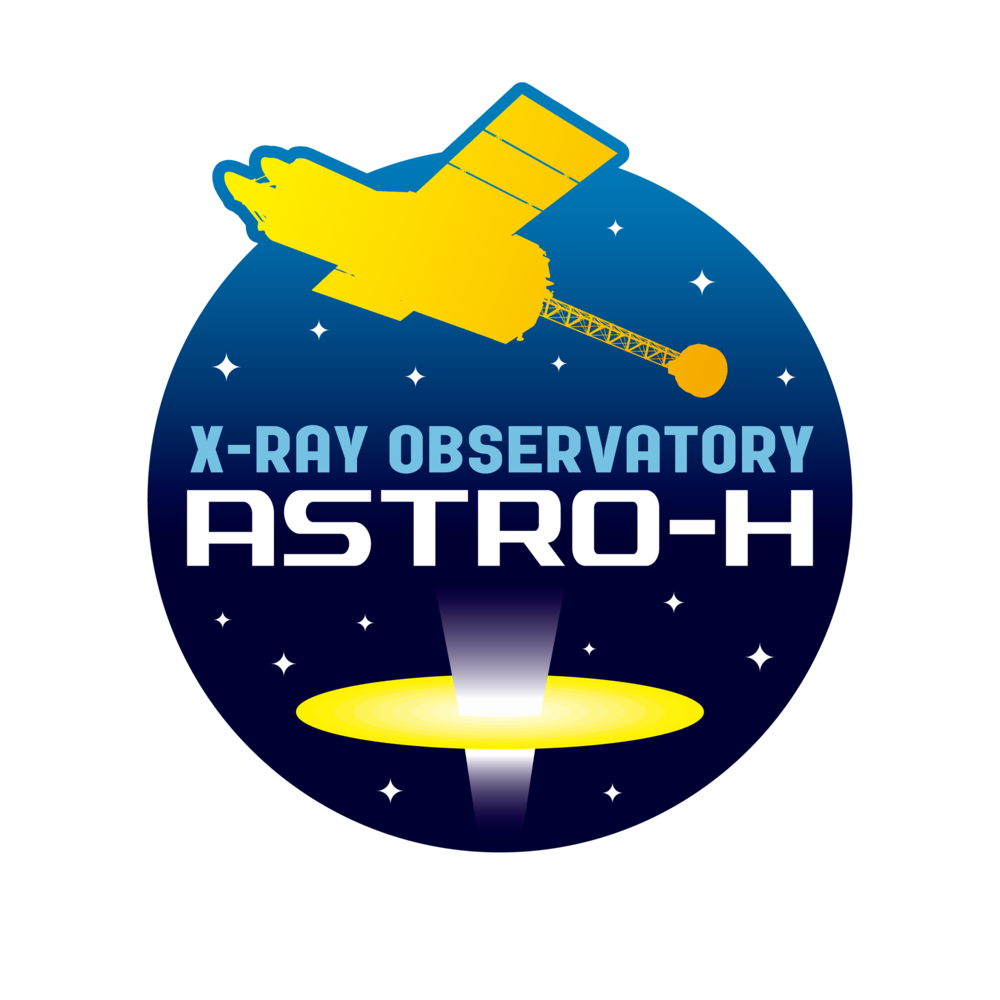}
			\end{center}
		\end{figure}
		\vspace{1cm}
		{\LARGE
		ASTRO-H Space X-ray Observatory\\
		White Paper\\
		}
		\vspace{5mm}
		{\large
		\WhitePaperTitle\\
		}
		\vspace{1cm}
		{
		\WhitePaperAuthors\\
		on behalf of the ASTRO-H Science Working Group
		}
	\end{center}
}

\usepackage{authblk}
\author[a]{Tadayuki~Takahashi}
\author[a]{Kazuhisa~Mitsuda}
\author[b]{Richard~Kelley}
\author[c]{Felix~Aharonian}
\author[d]{Hiroki~Akamatsu}
\author[e]{Fumie~Akimoto}
\author[f]{Steve~Allen}
\author[g]{Naohisa~Anabuki}
\author[b]{Lorella~Angelini}
\author[h]{Keith~Arnaud}
% ...
\affil[a]{Institute of Space and Astronautical Science (ISAS), Japan Aerospace Exploration Agency (JAXA), Kanagawa 252-5210, Japan}
\affil[b]{NASA/Goddard Space Flight Center, MD 20771, USA}
\affil[c]{Astronomy and Astrophysics Section, Dublin Institute for Advanced Studies, Dublin 2, Ireland}
\affil[d]{SRON Netherlands Institute for Space Research, Utrecht, The Netherlands}
\affil[e]{Department of Physics, Nagoya University, Aichi 338-8570, Japan}
\affil[f]{Kavli Institute for Particle Astrophysics and Cosmology, Stanford University, CA 94305, USA}
\affil[g]{Department of Earth and Space Science, Osaka University, Osaka 560-0043, Japan}
\affil[h]{Department of Astronomy, University of Maryland, MD 20742, USA}
% ...

\begin{document}

\newcommand{\WhitePaperTitle}{White Dwarf}
\newcommand{\WhitePaperAuthors}{
	K.~Mukai (NASA/GSFC/CRESST \& UMBC), T.~Yuasa (RIKEN), \\
	A.~Harayama (JAXA), 
	T.~Hayashi (JAXA), M.~Ishida (JAXA), 
	K.~S.~Long (STScI), \\
	Y.~Terada (Saitama University) and M.~Tsujimoto (JAXA)
}
\MakeWhitePaperTitle

\begin{abstract}
Interacting binaries in which a white dwarf accretes material
from a companion --- cataclysmic variables (CVs) in which the mass
loss is via Roche-lobe overflow, and symbiotic stars in which the white
dwarf captures the wind of a late type giant --- are relatively commonplace.
They display a wide range of behaviors in the optical, X-rays, and other
wavelengths, which still often baffles observers and theorists alike.
They are likely to be a significant contributor to the Galactic ridge
X-ray emission, and the possibility that some CVs or symbiotic stars
may be the progenitors of some of the Type Ia supernovae deserves
serious consideration.  Furthermore, these binaries serve as excellent
laboratories in which to study physics of X-ray emission from high
density plasma, accretion physics, reflection, and particle acceleration.
{\it ASTRO-H\/} is well-matched to the study of X-ray emission from
many of these objects.  In particular, the excellent spectral resolution
of the SXS will enable dynamical studies of the X-ray emitting plasma.
We also discuss the possibility of identifying an accreting,
near-Chandrasekhar-mass white dwarf by measuring the gravitational
redshift of the 6.4~keV line.
\end{abstract}
\clearpage

\maketitle
\clearpage

\tableofcontents
\clearpage

\section{Top Science}

The highest priority science of {\it ASTRO-H} for white dwarfs is
the search for extremely massive white dwarfs with high accretion rates.
We propose to do so by measuring the gravitational redshift of the
6.4 keV fluorescent Fe line from the white dwarf surface using the SXS.

Observationally speaking, this is a relatively simple experiment,
as long as a precise and accurate gain calibration is available for
the SXS. There are potential targets which are known to have a strong
enough 6.4 keV line and is thought to harbor a near Chandrasekhar mass
white dwarf. Unlike some other worthy investigations regarding accreting
white dwarfs, this is a study that requires X-ray spectroscopy, while
requiring relatively little multiwavelength support. Finally, if we can
determine just one white dwarf mass above 1.3 M$_\odot$ sufficient
accurately, that would be a major result with implications on the debate
over Type Ia supernova progenitor channels.


\section{Introduction}

Some white dwarfs in CVs (and possibly also symbiotic stars) are
magnetic enough that accretion proceeds along field lines.  The accretion
flow in such cases is nearly vertical with respect to the white dwarf
surface.  A strong stand-off shock forms just above the white dwarf surface;
the post-shock plasma must cool and further slow down before settling
onto the white dwarf.  This cooling often happens primarily via emission of
optically thin, thermal X-rays.  For a 0.6~$M_\odot$ white dwarf, the free-fall
velocity is $\sim$4,300 km\,s$^{-1}$ and the shock temperature is $\sim$22
keV (6,900 km\,s$^{-1}$ and 57 keV for a 1.0~$M_\odot$ white dwarf). 
For a specific accretion rate of 1 g\,s$^{-1}$cm$^{-2}$, the
immediate post-shock density is $\sim 10^{15}$ cm$^{-3}$
($\sim 6 \times 10^{15}$ cm$^{-3}$), the post-shock cooling timescale is 0.7 s 
(1.8 s), and the shock height is 5\% of the white dwarf radius, or
0.05~$R_\mathrm{WD}$ (0.34~$R_\mathrm{WD}$).  This basic picture of the accretion column
has been known for 40 years \citep{aizu1973}.  The observed X-ray emission
should be the sum of emission at many temperatures, as the plasma cools,
slows down, and becomes denser.  We explore below whether we can obtain
direct quantitative observational confirmation of this picture using 
velocity and density diagnostics of the high spectral resolution SXS data.

In non-magnetic CVs, the accretion proceeds via a disk.  This is probably
also true of the majority of symbiotic stars that have been detected above
$\sim$2 keV to date.  In such cases, the X-rays are emitted from the boundary
layer between the Keplerian disk and the white dwarf surface; the physics
of the boundary layer is far more complex than that of the accretion column.
Can we apply similar diagnostics as for magnetic CVs to aide the theoretical
efforts to understand the boundary layer?  Moreover, while our understanding
of the steady-state accretion disk is fairly secure, it is less so for dwarf
novae, for which the disk instability model (DIM) is widely adopted as the
explanation.  However, the basic version of DIM predicts the matter transferred
from the secondary to pile up in the disk during quiescence (the low state)
and hence very little accretion to take place onto the white dwarf.  The
observed X-ray luminosity is much higher than predicted.  One possible
modification of the DIM is that the quiescent disk has a central hole,
replaced by an advective flow. We will explore if {\it ASTRO-H} can
constrain the reflection amplitude with sufficient accuracy to determine
if such a hole exists in the quiescent disk.

If we can identify even a single massive white dwarf, close to the
Chandrasekhar limit, in an accreting binary, such a discovery have
a profound implications.  The presence of such a binary is an important
precondition for the single degenerate channel of Type Ia supernovae.
Conventional methods such as optical radial velocity studies have not
led to a secure identification of such a system.  We believe that
the {\it ASTRO-H} SXS can measure the gravitational redshift of
the 6.4 keV line produced via reflection on the white dwarf surface,
if it is sufficiently massive.  Although success is not guaranteed,
we consider this to be the most important {\it ASTRO-H} science
topic for accreting white dwarf binaries.

The 2--10 keV luminosity of CVs and symbiotic stars range from
$\sim 10^{29}$ ergs\,s$^{-1}$ for the low accretion rate dwarf novae
\citep{reisetal2013} to $> 10^{34}$ ergs\,s$^{-1}$ for the {\it Swift}
BAT detected symbiotic stars \citep{kenneaetal2009}.  Given their relative
proximity (of order 100 pc for many CVs and of order 2~kpc for numerous
symbiotic stars), the potential target list numbers several dozen.

\subsection{Abbreviations}
\begin{description}
	\item[BL] Boundary Layer via which gas from an accretion disk settles on to the white dwarf surface
	\item[CV] Cataclysmic Variable containing mass-accreting white dwarf
	\item[DIM] Disk Instability Model
	\item[PSR] Post-Shock Region of an accretion column
	\item[TNR] Thermo-Nuclear Runaway
	\item[WD] White Dwarf star
\end{description}

\section{Gravitationally redshifted 6.4 keV line in massive accreting WDs}

\subsection{Background and Previous Studies}

For the single degenerate channel for Type Ia supernovae to be viable,
there must exist accreting binaries hosting massive (near Chandrasekhar
mass) WDs.  The WD mass in CVs and symbiotic stars would show a secular
increase if accretion was the only factor involved.  However, under many
conditions, accreting WDs undergo thermonuclear runaways (TNRs) whenever a
sufficient mass is accumulated.  Such events are observed as nova eruptions,
in which a large amount of mass is ejected from the WD surface.  Observations
often show the nova ejecta to be enriched by the underlying WD material.  Given
this, it is unclear if WDs can grow in mass through successive accretion-TNR
cycles, and if so, under what conditions and at what efficiency (fraction
of the accreted mass that is retained by the WD).  This is a significant
weakness of the single degenerate channel.

Given this, even a discovery of single system with, say, $>1.3~M_\odot$
WD would be significant.  Such a binary was likely born with a somewhat
less massive WD, with the WD mass growing over time; if the WD mass is
decreasing, the initial mass of the WD must be even closer to the
Chandrasekhar limit, which seems an unlikely possibility.  Such a discovery
would prove that one of the necessary conditions for the single degenerate
channel to be viable is met.

There are indirect indications for high WD masses in accreting binaries.
For example, recurrent novae --- accreting WDs that have been seen to
experience nova eruptions multiple times over the last century or so
--- are likely to have high mass WDs.  This is because the critical density
required for a TNR is unlikely to be reached within a span of a few decades
after the previous TNR, unless the gravitational field of the WD is
exceptionally strong.  However, this line of reasoning is qualitative and
not foolproof.

Direct observational determinations of WD mass in accreting binaries are
often very imprecise.  Dynamical determination relies on radial velocity
studies, but few accreting WD systems are double-line spectroscopic
binaries --- the mass donors in CVs are often too faint, and those in
symbiotic stars are too bright.  Also, the radial velocity motion of the
WD is usually inferred via the motion of accretion disk around it; any
asymmetries or azimuthal structures can easily mislead us.  Furthermore,
the binary inclination is uncertain unless the system is eclipsing, which
often translates to large uncertainties in the derived WD mass.  

If we can measure the gravitational redshift of spectral features from
the white dwarf surface, this can all change.  We believe there is a
strong possibility that we can do so, by using {\it ASTRO-H} SXS to
measure the energy of the 6.4 keV line precisely.

\subsection{Prospects \& Strategy}

\begin{figure}[htb]
\begin{center}
\includegraphics[width=12cm]{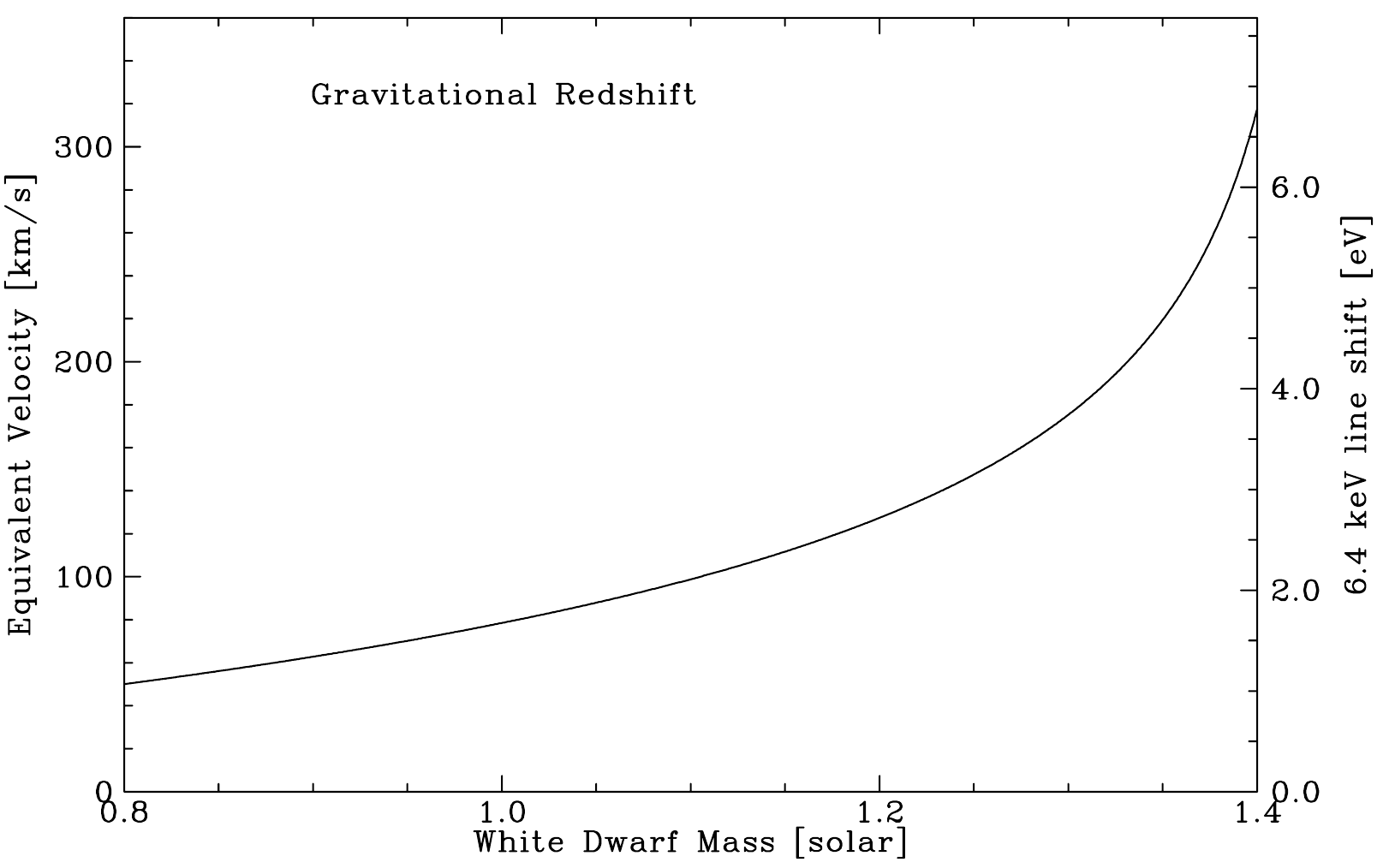}
\caption{Gravitational redshift as a function of the WD mass, expressed in
equivalent velocity and as energy shift for the 6.4 keV line.}
\label{figure:redshift}
\end{center}
\end{figure}

For this method to work, several conditions must be met.  The target must
be hard X-ray bright, with a significant 6.4 keV line as seen in previous
observations.  A large fraction of the 6.4 keV line flux must originate
on the WD surface.  The systemic velocity of the binary must be known,
and ideally the orbital motion of the WD negligible ($\sim$10 km\,s$^{-1}$).
Most importantly, the WD must be sufficiently massive that the gravitational
redshift can be measured against statistical and systematic uncertainties.

We show the expected degree of gravitational redshift
in Figure\,\ref{figure:redshift}.  As can be seen, the more massive the
WD, the easier it will be to detect the redshift.  For WD mass above
1.1 $M_\odot$, the redshift will be larger than 2 eV, which we argue
is easily within reach of SXS measurements.  Moreover, the gravitational
redshift is a sensitive proxy of the WD mass just below the Chandrasekhar
limit, precisely where a mass determination can have a significant
impact.

The first criterion can be satisfied by selecting a target detected
in the {\it Swift} BAT survey of the hard X-ray sky.  Several dozen
accreting WD systems have been detected to date, but most are magnetic CVs
with only moderately massive WDs ($0.8-1.0 M_\odot$).  Here we concentrate
on the 4 symbiotic stars detected in the BAT survey \citep{kenneaetal2009}.
Since the M giant mass donor is the dominant source of optical and IR
photons, their systemic velocity is (or can be) known accurately.
Due to the large orbit and the long orbital period, the radial velocity
motion of the WD is relatively small.  Finally, there is no evidence that
the WDs in these symbiotic stars are magnetic.  Therefore, the very
fact that they are detected by BAT indicates that the WDs are massive
(note that, for a given WD mass, the maximum temperature of the shock
is factor of $\sim$2 higher in the magnetic than in the non-magnetic
case, due to the difference in free-fall and Keplerian velocities).

\subsection{Targets \& Feasibility}

\begin{figure}[htb]
\begin{center}
\includegraphics[width=12cm]{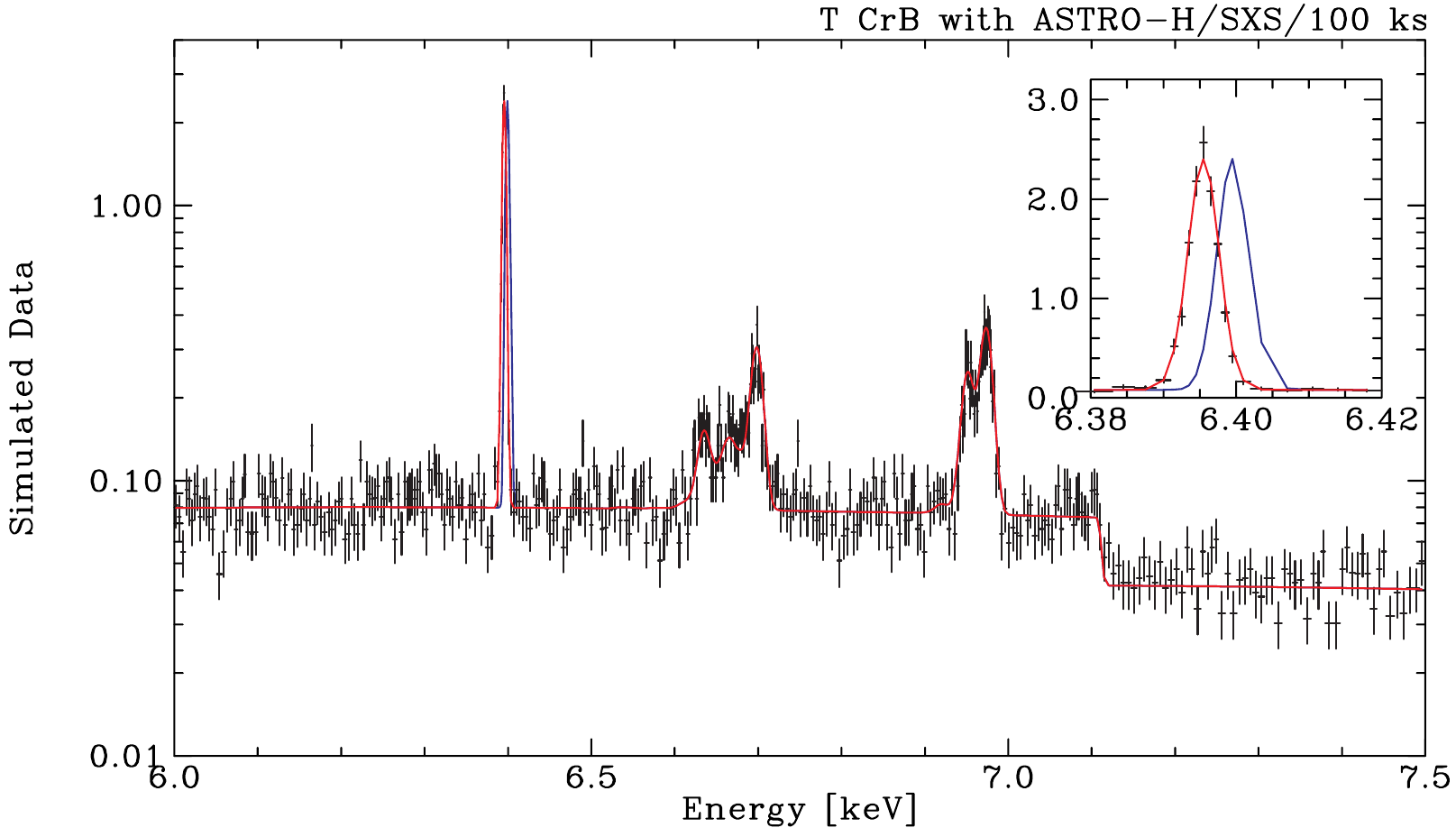}
\caption{Simulated {\it ASTRO-H} SXS 100 ks observation of T CrB in the Fe K$\alpha$ region.}
\label{figure:TCrB_sxs}
\end{center}
\end{figure}

One of the 4 symbiotic stars detected by BAT is T CrB, which is also
a recurrent nova.  Both this fact and the hard BAT spectrum indicates
that its WD is exceptionally massive, perhaps around 1.35 $M_\odot$.
It has been always active and hard X-ray bright during the BAT survey.
The WD photosphere is hidden in the optical/UV by the bright emissions from
the M giant mass donor and the accretion disk.   T CrB exhibits a strong
6.4 keV line, partly from the reflection off the WD surface.  We therefore
believe that T~CrB is the best target for this study.

We guide our simulation (Figure\,\ref{figure:TCrB_sxs}) using the 46~ks
{\it Suzaku} observation obtained in 2006.  The XIS data are fit using
a mildly broadened {\tt mkcflow} model absorbed by a partial covering
absorber with a single, narrow Gaussian at 6.396 keV.  The inset shows
the difference between this simulation and one with a narrow Gaussian at
6.400~keV.  This clearly shows that the {\it ASTRO-H} SXS has the
statistical quality necessary to detect a gravitational redshift of
4~eV. In the simulation, the following SXS instrumental response files were used:
\begin{itemize}
\item \verb|ah_sxs_5ev_basefilt_20100712.rmf| (energy redistribution file),
\item \verb|sxt-s_120210_ts02um_intallpxl.arf|(auxiliary response file), and 
\item \verb|sxs_cxb+nxb_5eV_20110211_1Gs.pha| (background spectral file).
\end{itemize}

We need to consider two further factors.  One is the accuracy of gain
calibration.  On this point, we propose to observe T~CrB with the calibration
source on to obtain the best possible gain calibration.  With this,
it is likely that the gain calibration is accurate enough for our
purpose.  However, we will pay close attention to this issue during
ground calibration and revisit our observing plan if necessary.

Another complication is that the intrinsic energy of the fluorescent
line is not fixed at 6.400 keV, and is in fact a function of the
ionization state of Fe.  The ionization state of Fe in the relevant
layer of the WD atmosphere is unknown.  Reflection requires high
column density of order 10$^{24}$~cm$^{-2}$, i.e., at or near the WD
photosphere.  If the white dwarf is hot, that alone (without irradiation
from above) can ionize Fe in the photosphere.  For example, lines of
Fe VII and Fe VIII are often seen in the photospheric spectra of 
PG~1159 type stars (hot, hydrogen deficient post-AGB stars), and up
to Fe X in the hottest cases such as PG~1159$-$035 ($T_\mathrm{eff}$=140,000K,
$\log g$=7; \citealt{werneretal2011}).  Since the photospheric densities
are high (of order $3\times10^{-5}$~g~cm$^{-3}$), and the hard X-ray
luminosity is modest ($\sim 10^{34}$~ergs\,s$^{-1}$), photoionization is
unlikely to results in ionization states higher than these.  Nevertheless,
the Fe K$\alpha$ energies change from 6.4055/6.3917~keV for Fe II to
6.4029/6.3900~keV for Fe VIII \citep{palmerietal2003}.  However, the
Fe K$\beta$ line energy changes much more significantly, and in the
opposite direction: 7.0583~keV for Fe II and 7.0740~keV for Fe VIII.
Thus, we will design our observation to be able to detect the Fe~K$\beta$
line, roughly 15\% the flux of the Fe~K$\alpha$ line.  If we can detect
both, the ionization state of Fe and the gravitational redshift can
both be determined from SXS data.

\subsection{Beyond Feasibility}

In addition, we can measure the maximum temperature in the shock,
$kT_{max}$ using the HXI data.  In a 100 ks observation, we estimate
a statistical accuracy of about $\pm$2 keV.  Actual accuracy depends
on the parameter degeneracy between $kT_{max}$ and the reflection amplitude,
background subtraction accuracy and other systematics.  This will provide
a good cross-check for the gravitational redshift measurement.

The spectrum of T~CrB is highly absorbed, and the nature of intrinsic
absorber in symbiotic stars is poorly understood.  The proposed observation
will provide the most precise determination yet of the absorber, including
any variability during the observation.

Finally, we will also obtain high quality spectra of the He-like and H-like
lines of Fe.  We will extract dynamical information regarding the post-shock
region from these lines, as we propose to do for SS~Cyg.  If the emission
region is sufficiently close to the white dwarf, the H-like lines will also
be gravitationally redshifted.  If that is the case, it is in some way
a cleaner measurement than the reflection line, since the ionization state
of the line-producing ions is not in doubt in this case.

\section{Detailed structure of the post-shock region in magnetic CVs}

\subsection{Background and Previous Studies}

\begin{figure}[htb]
	\begin{center}
		\begin{minipage}{0.48\hsize}
			\begin{center}
			\includegraphics[width=0.95\hsize, bb= 9 9 966 641]{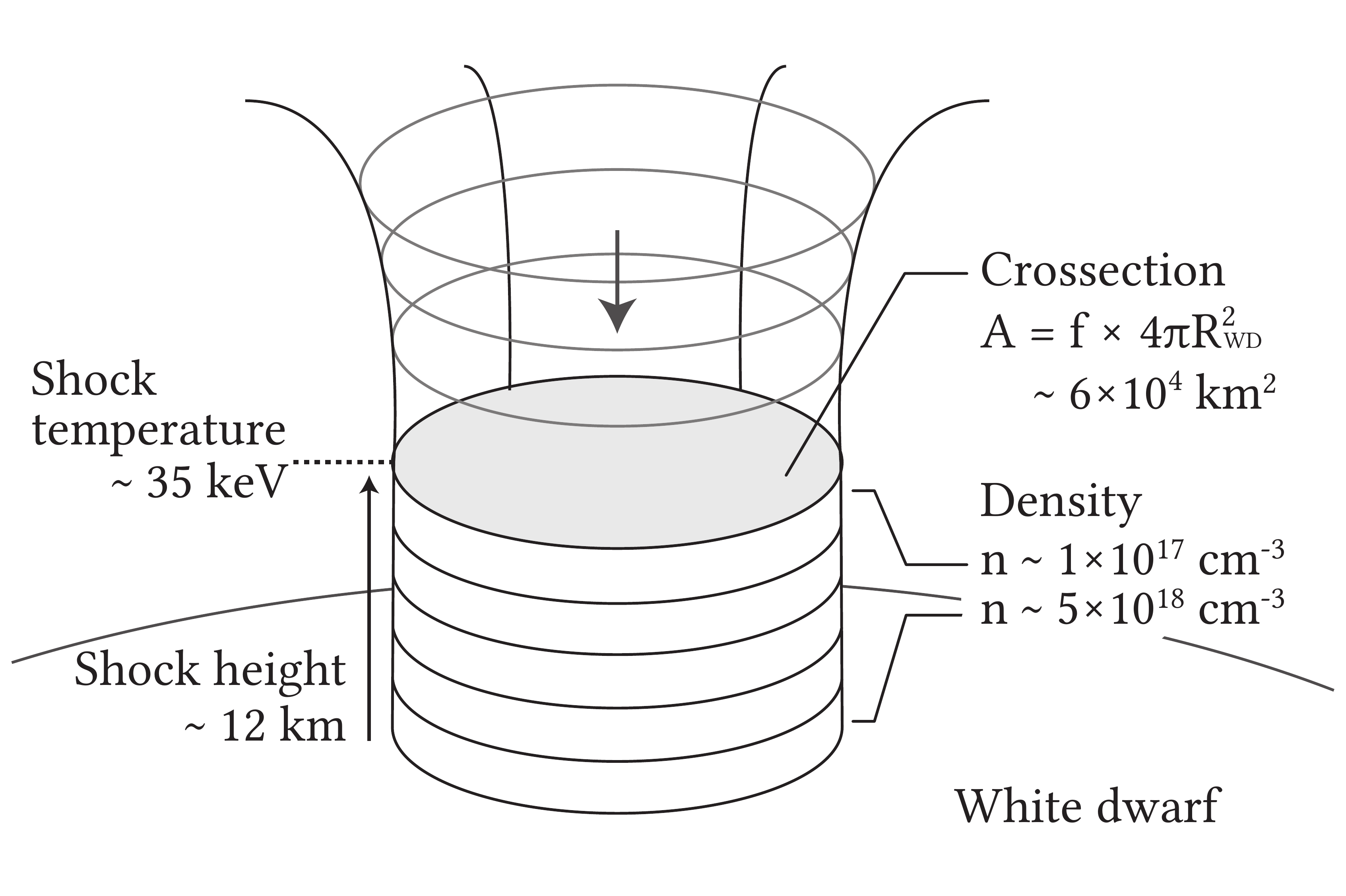}
			\caption{Schematic view of an accretion column and a post-shock region of an intermediate polar). Values presented in the figure are calculated using typical accretion condition of V1223 Sgr; $M_{\mathrm{WD}}=0.79~M_\odot$, $f=0.001$ and $\mathrm{d}M/\mathrm{d}t=8.4\times10^{16}~\mathrm{g}~\mathrm{s}^{-1}$ \citep{yuasaetal2010,hayashietal2011}.}
			\label{figure:ip_accretion_column_wireframe}
			\end{center}
		\end{minipage}
		\hspace{0.02\hsize}
		\begin{minipage}{0.48\hsize}
			\begin{center}
			\includegraphics[width=0.95\hsize, bb= 0 0 585 516]{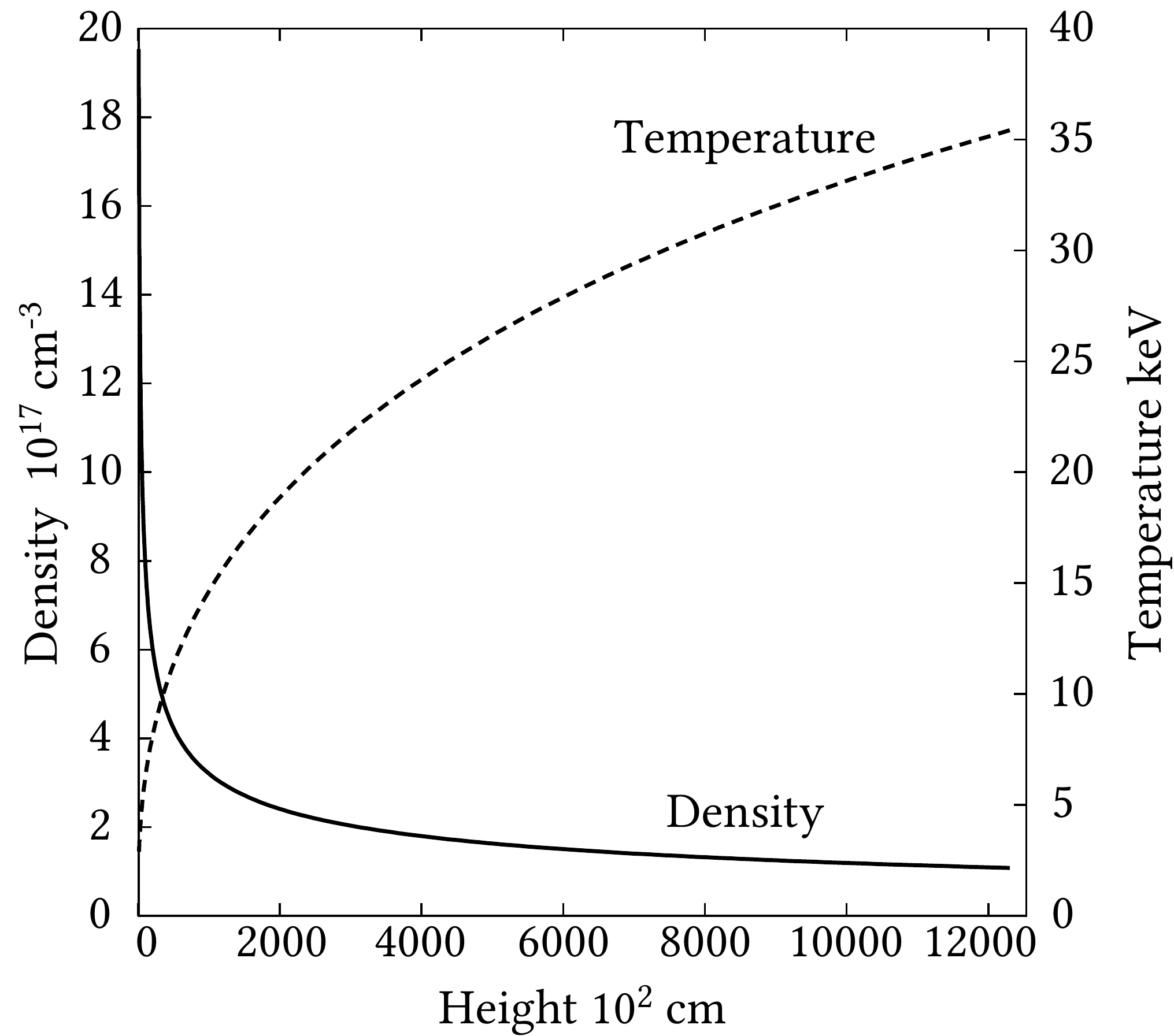}
			\caption{Plasma density and plasma temperature profiles in the post-shock region calculated using the numerical hydrostatic model of \citet{yuasaetal2010}. Assumed accretion parameters are the same as presented in Figure \ref{figure:ip_accretion_column_wireframe}.}
			\label{figure:plot_mwd079f0001_n_kT}
			\end{center}
		\end{minipage}
	\end{center}
\end{figure}

As mentioned previously, the basic picture of the post-shock region (PSR)
of magnetic CVs is relatively secure. However, the specific accretion
rate (or, conversely, the fractional area of the white dwarf surface
onto which accretion occurs) is a key parameter that is poorly known.
It is determined by the complex interaction between the accretion flow
and the magnetic field, which is difficult to solve purely theoretically.
In practice, we can treat the specific accretion rate (usually assumed to
be uniform across the accretion column) as a free parameter, then solve
for the temperature, velocity, and density structure of the PSR, and then
predict the resulting multi-temperature X-ray spectrum.  We can then fit
such models to the observed X-ray spectrum to obtain, e.g., the white
dwarf mass \citep{yuasaetal2010}. Figures \ref{figure:ip_accretion_column_wireframe} and \ref{figure:plot_mwd079f0001_n_kT} present
a schematic view of a PSR, and post-shock plasma density/temperature
profiles obtained by solving numerical hydrostatic equations shown in \citet{yuasaetal2010}.
In the figures, typical parameters are set assuming those of V1223 Sagittarii,
one of the classical magnetic CVs that has relatively high accretion rate.

For most magnetic CVs, the PSR density is expected to be so high that
He-like triplets that can be resolved with {\it Chandra} HETG do not
help.  In addition, the HETG has a small effective area, so it is
not practical for most magnetic CVs.  The only exception is a highly
atypical magnetic CV, EX Hya, for which some density information has
been obtained using Fe L density diagnostics \citep{maucheetal2003}.

The {\it ASTRO-H} SXS will be able to resolve Fe XXV triplets into
separate lines providing considerably larger effective area at the same
time. Because of this, we aim to perform density diagnostics based on
Fe XXV triplets and constrain the geometry of a PSR for selected magnetic
CVs. Determination of the
geometry will lead to precise calculation of density, temperature, and velocity
distribution along the PSR, and further understanding on accretion physics
and magnetic field structure of CVs.

For review of density and temperature diagnostic methods using triplets from
He-like ions, see \citet{porquetetal2010}.

\subsection{Prospects \& Strategy}
\subsubsection{Density diagnostics with Fe XXV triplets}
\begin{figure}[htb]
	\begin{center}
		\begin{minipage}{0.55\hsize}
			\begin{center}
			\includegraphics[height=6.5cm, bb= 14 9 317 225]{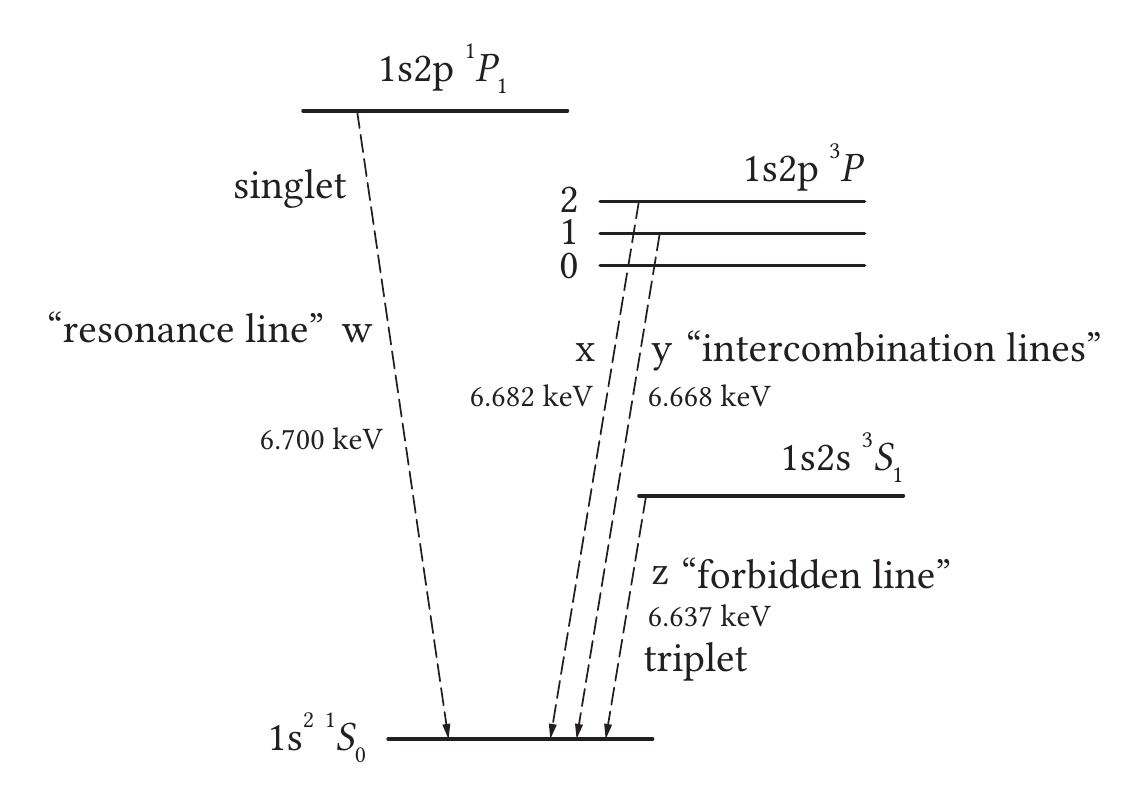}
			\caption{Radiative transitions of Fe XXV ions relevant to the present magnetic CV studies. See \citet{gabrieljordan1969,porquetetal2010} for details.}
			\label{figure:fe_xxv_triplet_levels}
			\end{center}
		\end{minipage}
		\hspace{0.01\hsize}
		\begin{minipage}{0.42\hsize}
			\begin{center}
			\includegraphics[width=0.98\hsize, bb = 0 0 498 503]{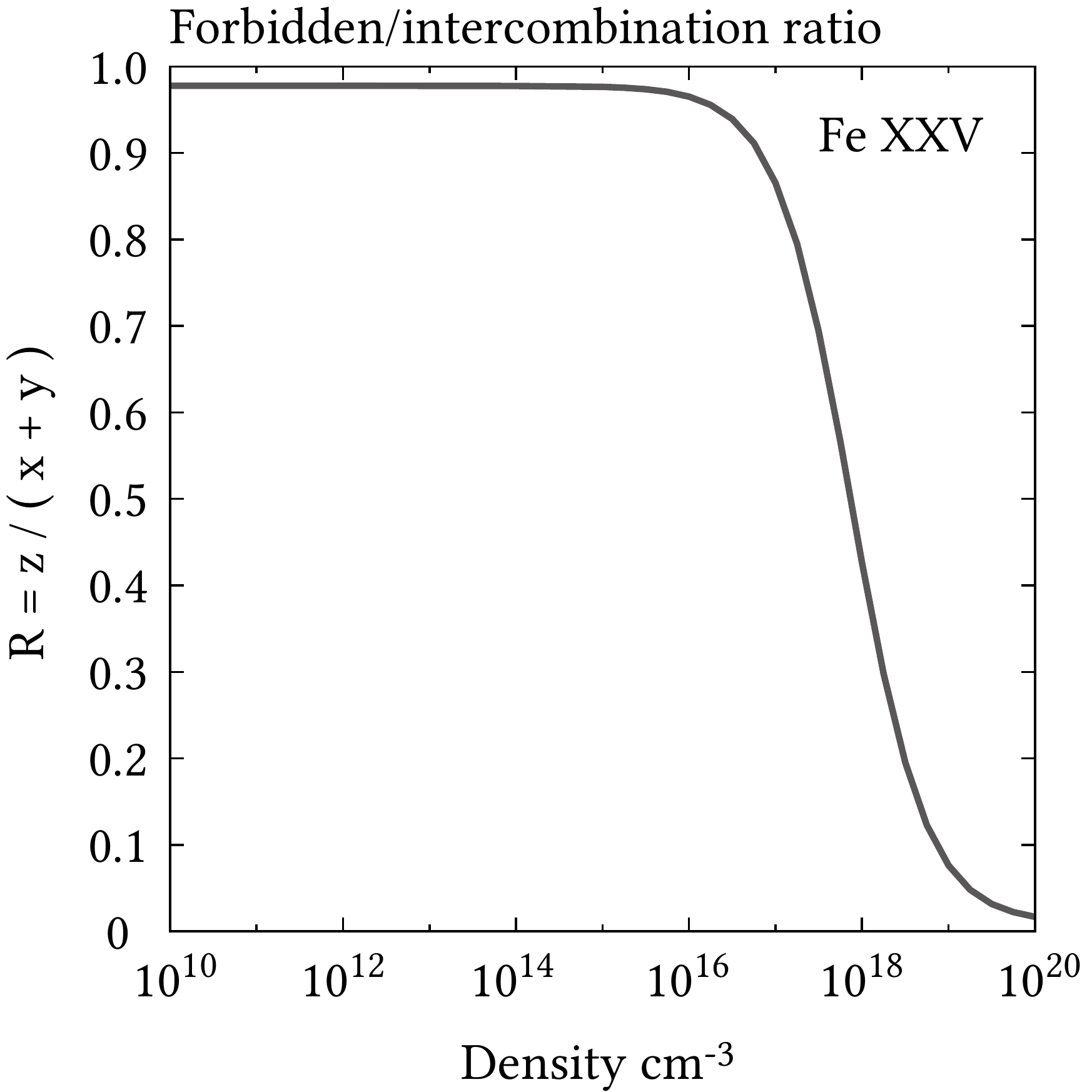}
			\caption{He-like $R=z/(x+y)=f/i$ line ratio of Fe.}
			\label{figure:plot_r_fe_he}
			\end{center}
		\end{minipage}
	\end{center}
\end{figure}

Figure \ref{figure:fe_xxv_triplet_levels} shows radiative transitions of Fe XXV ions.
The intensity of the forbidden line ($^3S_1$-$^1S_0$) is strongly dependent on the
plasma density.  Above a ``critical density'' collisional de-excitation from $^3S_1$ to $^1S_0$
becomes the dominant process and the ``forbidden'' radiative decay is suppressed.
The forbidden-intercombination line intensity ratio $R\equiv z/(x+y)$, where $x$, $y$ and $z$ are
intensities of corresponding lines, can therefore be used to measure the density of originating plasma
based on observed spectra.
Figure \ref{figure:plot_r_fe_he} illustrates this dependence based on triplet line intensities calculated
with the collisional ionization equilibrium (CIE) model available in the SPEX analysis package \citep{spex1996}.
Based on this curve, the critical density of Fe XXV triplet is $\sim10^{17}-10^{18}~\mathrm{cm}^{-3}$. 

Astrophysical plasmas with a density above the critical density of Fe XXV
together with plasma temperature that is high enough to ionize Fe to He-like
are rather rare.  However, PSRs of magnetic CVs with high accretion rates
are believed to reach this extreme condition as exemplified in Figures 
\ref{figure:ip_accretion_column_wireframe} and \ref{figure:plot_mwd079f0001_n_kT}.
Therefore, by studying the Fe XXV triplet lines with the SXS, plasma density in a PSR
could be directly measured, and it will be possible to constrain the physical condition
of a PSR further deepening our understanding on accretion physics.

\subsubsection{Spectral model of the PSR updated for the SXS observation}
To evaluate feasibility of the density diagnostics based on the Fe XXV triplet line ratio,
we updated the X-ray spectral model of the PSR \citep{yuasaetal2010}, using the CIE
model of the SPEX replacing the APEC CIE model. Essence of this model construction
is to convolve the emission measure distribution and single temperature CIE spectrum
over the whole PSR. By doing so, the new model is now able to include the density 
dependence of the forbidden/intercombination lines correctly accounting the multi-temperature
nature of the PSR plasma.

When calculating the spectral model, it is necessary to fix (or provide) a WD mass value
which is the primary free parameter that changes model spectral shape, and the geometrical
area of the PSR as the fraction of the WD surface area. This is because, for a fixed
total accretion rate, the density of a PSR depends on the area over which accretion occurs,
and so do solutions of the hydrostatic equations used in the model (i.e. different density/temperature
distributions can be obtained). In the present paper, we use $f$ parameter, which is defined
as a ratio of accretion column cross section to the WD surface area, to denote the geometrical
size of the PSR. Three example values, 0.0002, 0.001, and 0.005 are chosen from a possible
range of this parameter based on previous studies of magnetic CVs (e.g. \citealt{hellier1997}).

\begin{figure}[htb]
	\begin{center}
		\begin{minipage}{0.42\hsize}
			\begin{center}
			\includegraphics[width=\hsize, bb= 9 9 1321 840]{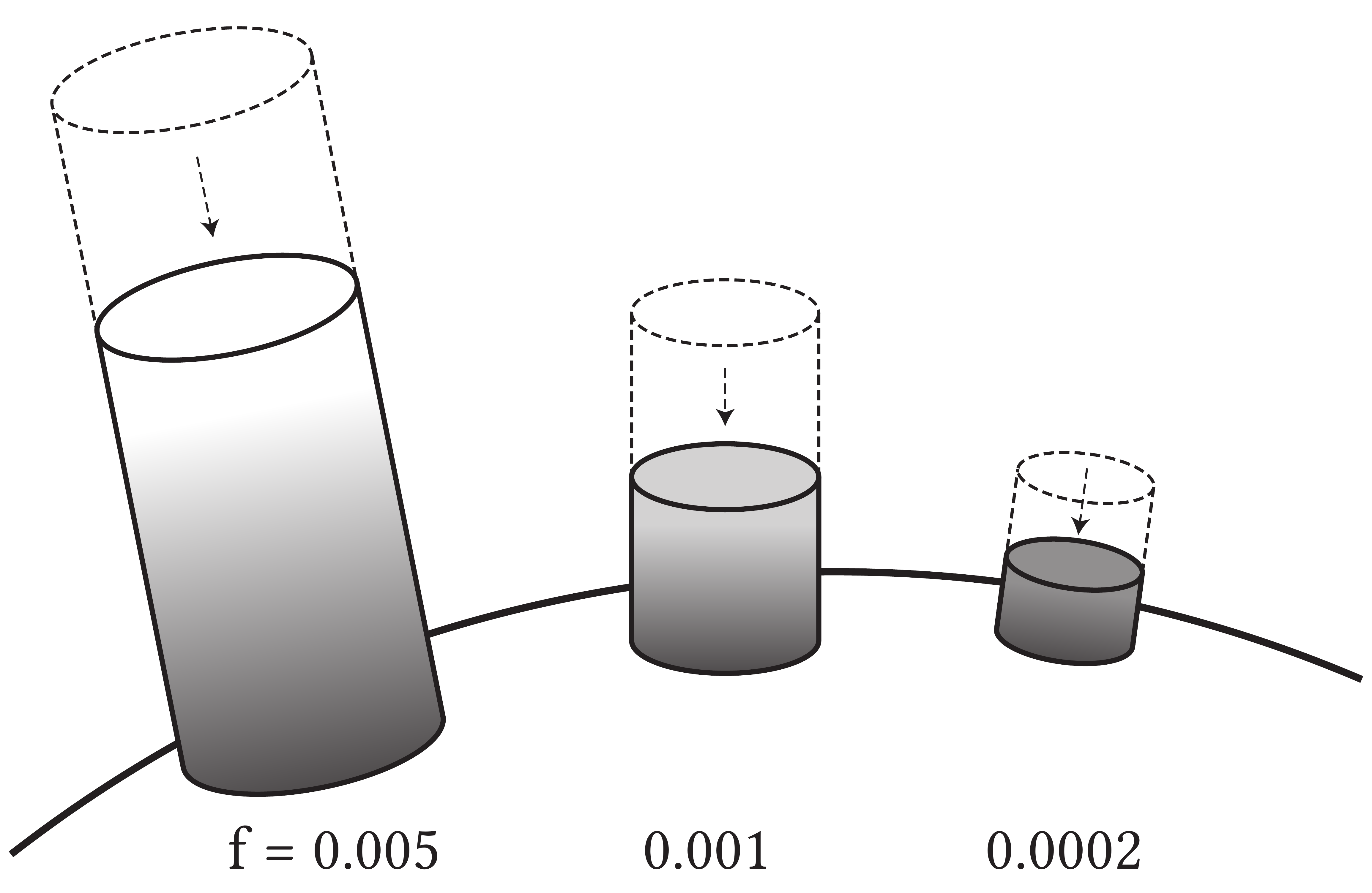}
			\caption{Schematic view of PSR structures with different cross sections but the same total mass accretion rates. The thicker parts correspond to higher densities. The scale of the PSR and the shock heights are exaggerated.}
			\label{figure:v1223_psr_for_different_f}
			\end{center}
		\end{minipage}
		\hspace{0.02\hsize}
		\begin{minipage}{0.53\hsize}
			\begin{center}
			\includegraphics[width=0.95\hsize, bb= 0 0 539 511]{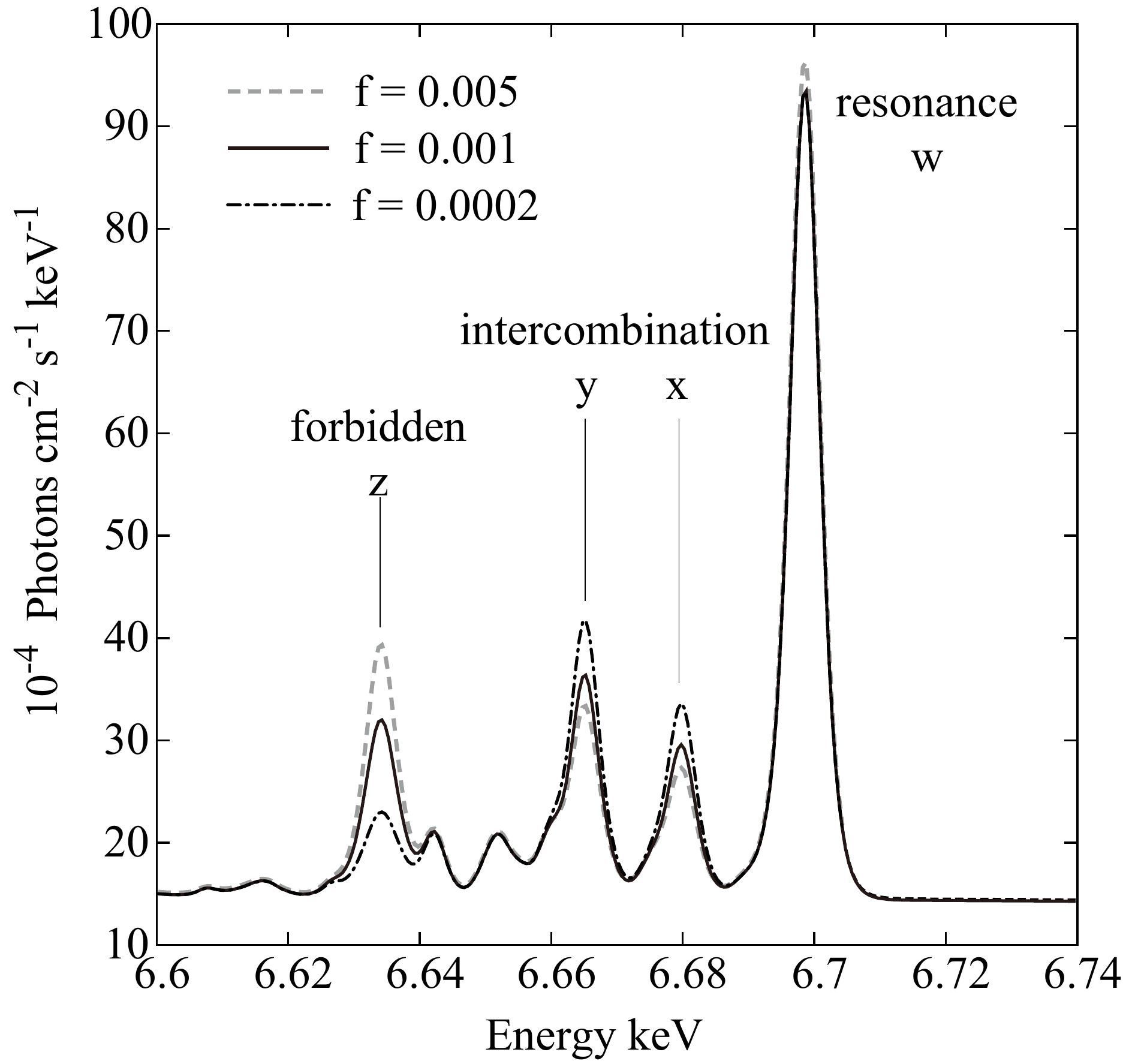}
			\caption{Close-up view of Fe XXV K$\alpha$ spectra of X-ray emission calculated for multi-temperature and multi-density post-shock plasma of V1223 Sgr. Dashed, solid, and dash-dotted curves are calculated with three different covering fraction $f$ of 0.005, 0.001, and 0.0002. Note that the solid curve corresponds to total spectrum expected from temperature/density the distribution presented in Figure \ref{figure:plot_mwd079f0001_n_kT}.}
			\label{figure:total_spectra_mwd079}
			\end{center}
		\end{minipage}
	\end{center}
\end{figure}

As illustrated in Figure \ref{figure:v1223_psr_for_different_f}, the hydrostatic equations of a PSR
with different values of $f$ result in different PSR structures, especially in terms of shock heights and plasma density below the
shock. Qualitatively, this can be understood that as the PSR cross section gets smaller, the plasma
density should increase leading to effective plasma cooling. With high densities, shock heated plasma
can cool down effectively, and satisfy the ``zero temperature'' boundary condition at the bottom of
the PSR even if it has higher shock temperature (i.e. smaller shock height) compared to PSRs with
larger $f$ values.

After fixing $f$ at the values listed above, we calculated total spectra emitted from assumed PSRs as presented in
Figure \ref{figure:plot_mwd079f0001_n_kT}. To apply the spectral model to SXS observation
simulation in the following sections, we used the estimated parameters of V1223 Sgr in the calculation
as listed in Table \ref{table:model_calculation_parameters_v1223sgr}. 
Table \ref{table:model_calculation_result_v1223sgr} 
presents calculation results of some representative values.
Note that, in the calculation, we also took into account redshift caused by the bulk motion, or falling
velocity, of plasma which amounts about $1500~\mathrm{km}~\mathrm{s}^{-1}$ at the top of a PSR
although the redshift is little apparent in the resulting spectra (this is because most of Fe XXV K$\alpha$
emission comes from regions where falling velocity is small, i.e. lower parts of a PSR).

It is apparent from Figure \ref{figure:plot_r_fe_he} that the forbidden line (labeled z) is
sensitive to $f$, i.e., as $f$ decreases (or plasma density below the shock increases) intensity
of the line decreases due to depopulation of $^3S_1$ level by collisional excitation (see above).
This confirms our previous prospect that the $R$ ratio can be used for density diagnostics
of PSRs of magnetic CVs with high accretion rates like V1223 Sgr.

\begin{table}[htb]
\begin{center}
\caption{Parameter values assumed in the PSR structure model calculation. }
\label{table:model_calculation_parameters_v1223sgr}
\small{
{\renewcommand\arraystretch{1.25}
\begin{tabular}{cclccc}
\hline
\multicolumn{2}{c}{Assumed values} & \multicolumn{1}{c}{Note}\\
\hline
$M_{\mathrm{WD}}$ & $0.79~M_{\odot}$ & \footnotesize{WD mass \citep{yuasa2013}.}\\
$R_{\mathrm{WD}}$ & 7070~km & \footnotesize{WD radius calculated for 0.79~$M_\odot$ WD using \citep{nauenberg1972}.}\\
$\mathrm{d}M/\mathrm{d}t$ & $8.4\times10^{16}~\mathrm{g}~\mathrm{s}^{-1}$ & \footnotesize{Total mass accretion rate \citep{hayashietal2011}.}\\
$Z$ & $0.29~Z_{\odot}$ & Metal abundance of accreting gas.\\
\hline
\end{tabular}
}
}
\end{center}
\end{table}

\begin{table}[htb]
\begin{minipage}{\hsize}
\begin{center}
\caption{Result of the PSR structure model calculation. }
\label{table:model_calculation_result_v1223sgr}
\small{
{\renewcommand\arraystretch{1.25}
\begin{threeparttable}
\begin{tabular}{cccc}
\hline
Result & \multicolumn{3}{c}{$f$}\\
 & 0.0002 & 0.001 & 0.005\\
 \hline
$h_{\mathrm{s}}$~cm & $2.52\times10^5~$ & $1.25\times10^6$ & $6.14\times10^6$\\
$kT_{\mathrm{s}}$~keV & 35.7 & 35.7 & 35.4\\
$\rho_{\mathrm{s}}$~cm$^{-3}$ & $5.42\times10^{17}$ & $1.08\times10^{17}$ & $2.18\times10^{16}$ \\
$v_{\mathrm{s}}$~cm~s$^{-1}$ & $1.36\times10^8$ & $1.36\times10^8$ & $1.36\times10^8$\\
\hline
\end{tabular}
\begin{tablenotes}\footnotesize
\item[*] $h_{\mathrm{s}}$, $kT_{\mathrm{s}}$, $\rho_{\mathrm{s}}$, and $v_{\mathrm{s}}$ are 
shock height measured from the WD surface, plasma temperature, density, and falling velocity
directly below the shock.
\end{tablenotes}
\end{threeparttable}
}
}
\end{center}
\end{minipage}
\end{table}

\subsection{Targets \& Feasibility}
\subsubsection{V1223 Sgr, the brightest magnetic CV, as an appropriate target}
\begin{figure}[htb]
\begin{center}
\includegraphics[width=\hsize, bb= 8 0 919 256]{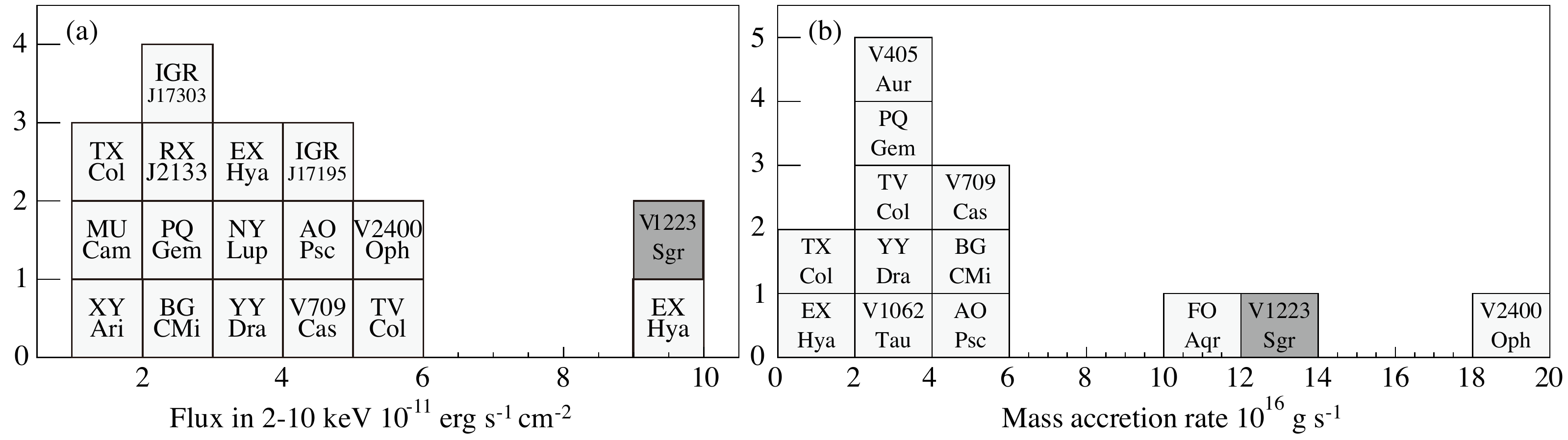}
\caption{Histograms of (a) $2-10$ keV flux and (b) mass accretion rate of well-studied classical magnetic CVs (intermediate polars). To highlight V1223 Sgr, bins with an entry for the source is labeled with source name. Data from \citet{yuasaetal2010} and \citet{suleimanovetal2005}.}
\label{figure:ip_histogram}
\end{center}
\end{figure}

To perform precise density diagnostics, it is essential to select targets that have
enough high X-ray flux for high counting statistics and accretion rates that
result PSR density higher than the critical density of Fe XXV triplet lines.
For selecting appropriate target based on the present knowledge, we created
histograms of $2-10$~keV flux and mass accretion rate as shown in Figure \ref{figure:ip_histogram}.
As can be easily seen in the histograms, V1223 Sgr is one of the brightest well
known magnetic CVs, and at the same time, it has a relatively high total mass accretion
that can create different forbidden line intensities depending on the PSR cross section $f$
(as presented in Figure \ref{figure:total_spectra_mwd079}).

Based on these facts, V1223 Sgr is an ideal target for an early observation
with {\it ASTRO-H}. Not only the density diagnostics, but studies of X-ray
reflection at the WD surface can be performed at the same time using a
single observation of V1223 Sgr (see next section).

\subsubsection{Simulation of SXS spectra}
Using the calculated spectral model (Figure \ref{figure:plot_mwd079f0001_n_kT}),
we simulated spectra obtained with the SXS using the following instrumental response
files:
\begin{itemize}
\item \verb|ah_sxs_5ev_basefilt_20100712.rmf| (energy redistribution file),
\item \verb|sxt-s_100208_ts02um_intallpxl.arf| (auxiliary response file), and 
\item \verb|sxs_nxb_5ev_20110211_1Gs.pha| (background spectral file).
\end{itemize}
We assumed a net exposure of 100 ks to achieve statistically sufficient photon 
counts for detailed analysis of w, x, y, and z line intensities. 
Figure \ref{figure:total_spectra_mwd079} presents 
three resulting spectra for $f=0.0002$, 0.001, and 0.005. Although we do not know
the actual $f$ value for V1223 Sgr, we believe that the three values cover most
of a possible range, and therefore. 

\begin{figure}[htb]
	\begin{center}
		\begin{minipage}{0.48\hsize}
			\begin{center}
			\includegraphics[height=10cm, bb= 0 0 692 921]{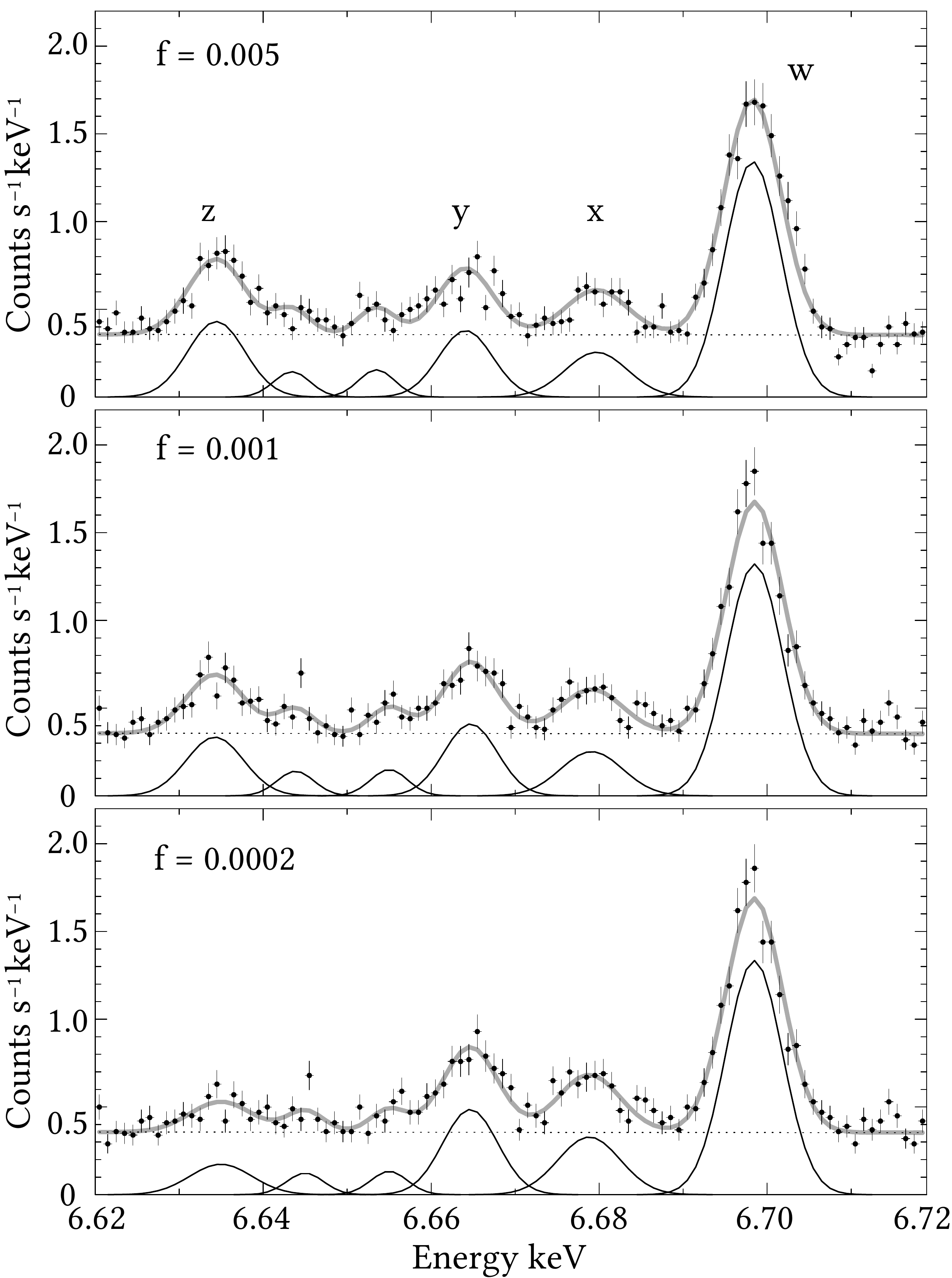}
			\caption{Fe XXV K$\alpha$ emission line spectra of the PSR of V1223 Sgr simulated assuming a 100-ks SXS observation. As used in Figure \ref{figure:total_spectra_mwd079}, we simulated three cases with different covering fractions $f=0.005$, 0.001, and 0.0002.}
			\label{figure:v1223_feheka_mwd079f00002_0001_0005_100ks}
			\end{center}
		\end{minipage}
		\hspace{0.02\hsize}
		\begin{minipage}{0.48\hsize}
			\begin{center}
			\includegraphics[height=7cm, bb=0 0 499 494]{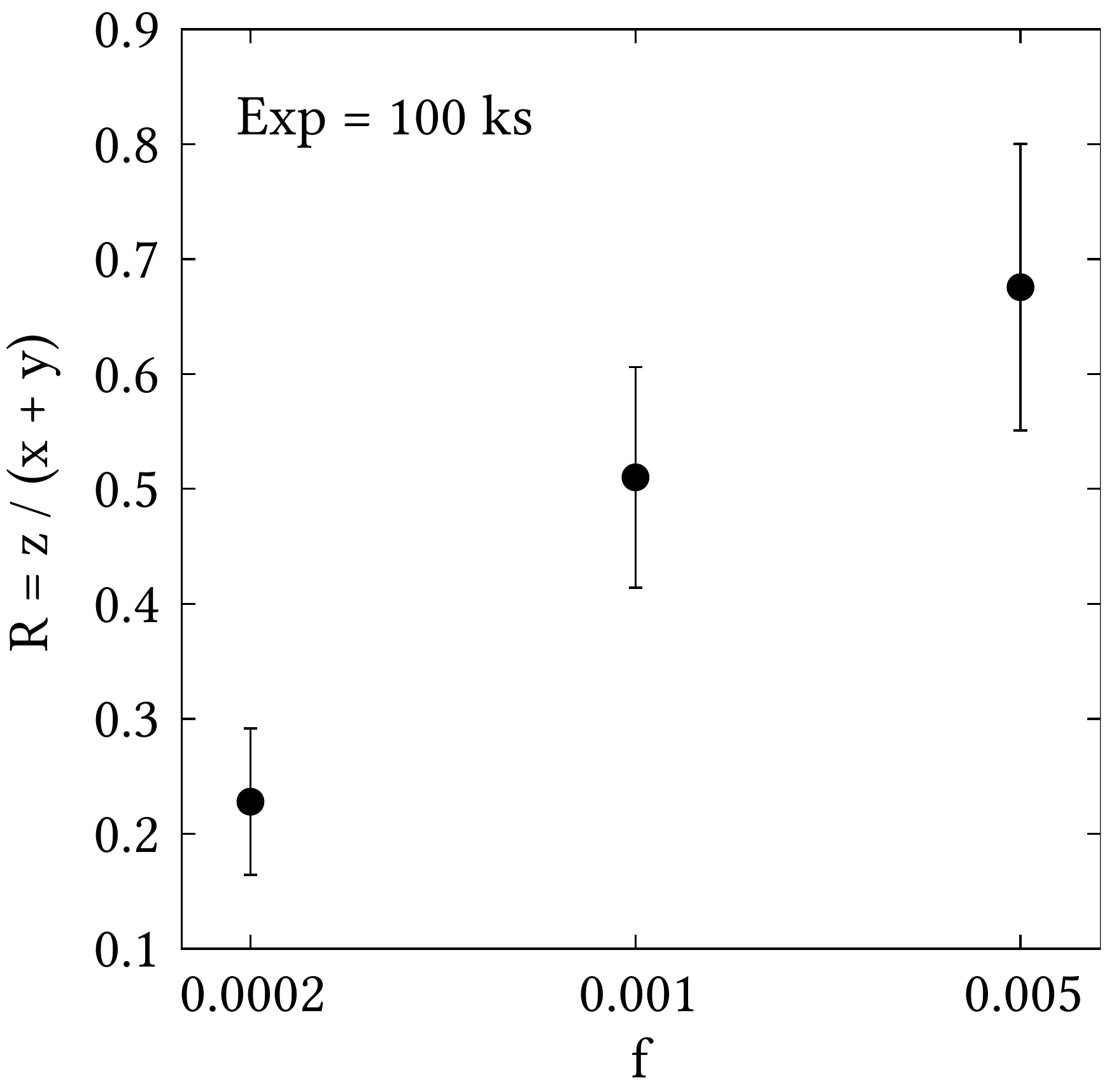}
			\caption{R ratios calculated from the x, y, and z intensities obtained from the gaussian fits of simulated 100-ks spectra. Associated errors are at the 90\% confidence level. }
			\label{figure:plot_RFactor_100_160ks}
			\end{center}
		\end{minipage}
	\end{center}
\end{figure}

In each spectrum, we fitted emission lines with
Gaussians, and estimated their intensities accompanied with errors at the 90\%
confidence level as results listed in Table \ref{table:v1223_gaussian_fit_result}.
From the fitting result, we calculated $R=z/(x+y)$, and plotted against $f$ in 
Figure \ref{figure:plot_RFactor_100_160ks}. Based on the calculated $R$ results,
we expect that we can clearly distinguish $f=0.0002$ and 0.001 cases, but for
0.001 and 0.005 cases, discrimination may be possible only marginally within the errors.

Although plasma density diagnostics have been a very important method which can apply to broad range of densities and wavelengths, those with Fe XXV triplet lines are only possible with the SXS, and in particular, with targets that exhibits high-rate mass accretion. By observing V1223 Sgr, we would like to exploit the SXS capability and realize one of long-standing challenges anticipated for X-ray micro calorimeters.

%

\begin{table}[htb]
\begin{center}
\caption{Results of the Gaussian fits to the simulated spectra of a 100-ks observation of V1223 Sgr. }
\label{table:v1223_gaussian_fit_result}
\small{
{\renewcommand\arraystretch{1.25}
\begin{tabular}{cccccc}
\hline
$f$ & \multicolumn{4}{c}{Intensity \footnotesize{($10^{-5}~\mathrm{photons}~\mathrm{s}^{-1}~\mathrm{cm}^{-2}$)}} & $R$\\
 & w & x & y & z & \\
\hline
0.005 & $4.80^{+0.25}_{-0.37}$ & $1.00^{+0.18}_{-0.30}$ & $1.29^{+0.11}_{-0.31}$ & $1.54^{+0.22}_{-0.19}$ & $0.68\pm0.12$ \\
0.001 & $4.72^{+0.30}_{-0.37}$  & $0.99^{+0.19}_{0.27}$ & $1.38^{+0.20}_{-0.19}$ & $1.20^{+0.11}_{-0.21}$ & $0.51 \pm0.10$ \\
0.0002 & $4.77^{+0.26}_{-0.33}$ & $1.29^{+0.22}_{-0.18}$ & $1.68^{+0.23}_{-0.18}$ & $0.67^{+0.17}_{-0.18}$ & $0.23 \pm0.06$ \\
\hline
\end{tabular}
}
}
\end{center}
\end{table}

\subsection{Beyond Feasibility}

The same observation of V1223~Sgr can be used to study reflection
as a function of spin phase, and to search for gravitationally redshifted
6.4 keV line (see below).  In addition, the spectrum of magnetic CVs
are affected by complex partial covering absorber \citep{DM1998},
including warm absorber features \citep{mukaietal2001}.  The SXS has
large enough effective area and adequate spectral resolution below 1 keV
to study these.

\section{Refection in accreting WDs}

\subsection{Background and Previous Studies}
The primary X-rays emitted by accreting WDs can reflect
off the WD surface and pre-shock accretion disk/column.
This results in the reflection bump in $>$10~keV continuum and
the 6.4 keV fluorescent line, and a physical understanding requires
both to be fit in a self-consistent manner.  An accurate characterization
of the reflection bump is necessary for an accurate determination
of the shock temperature, and hence the white dwarf mass. 
Unfortunately, most CVs and symbiotic stars are faint enough above 10~keV
that the systematic uncertainties of the background in non-imaging
detectors, including {\it Suzaku} HXD/PIN, have proved to be
the limiting factor.

\subsection{Prospects \& Strategy}
The combination of the high sensitivity of the {\it ASTRO-H} HXI for hard
continuum and the spectral resolution of the SXS will allow us to measure
the reflection fraction (both the hard continuum bump and the 6.4 keV
fluorescent line) with an unprecedented accuracy.  For the X-ray brightest
magnetic CVs, this can be done in several spin phases, which will test
angle-dependent models of reflection.  The 6.4 keV line may exhibit Compton
shoulder in some cases. As first reported by \cite{hayashietal2011} in V1223 Sgr,
fluorescence from free-falling pre-shock gas will have detectable redshift
due to its line-of-sight velocity of several$\times1000$~km~s$^{-1}$.
A detection of redshifted Fe fluorescent line will improve our understanding
of the accretion stream in magnetic CVs.  This can lead to further
improvements in the PSR model calculation and, consequently, spectral
model calculations.

Hard X-ray continuum spectra of bright magnetic CVs may be observed with
{\it NuSTAR} well before {\it  ASTRO-H}, and the combination of the
{\it NuSTAR} data with those of Fe K$\alpha$ lines from {\it XMM-Newton}
or {\it Suzaku} may help to some extent. However, {\it ASTRO-H} data
will definitely offer an improvement since they will allow a fully physical
treatment of geometry, fluorescent line from WD surface and pre-shock gas
and its Comptonization using simultaneous HXI and SXS data.

\subsection{Targets \& Feasibility}
The same accreting magnetic CV, V1223 Sgr, as the PSR structure
study is the most appropriate target because this is the brightest CV
in the hard X-ray band ($>$10~keV), and therefore, we can expect sufficient
photon counts within a realistic exposure. In addition, although this is a
well studied classical magnetic CV, there are discrepancies among published
analysis of the reflection component
 using instruments onboard past and 
current missions (e.g. \citealt{revnivtsev2004,hayashietal2011}).

\begin{figure}[htb]
\begin{center}
\includegraphics[width=12cm, bb= 0 0 512 268]{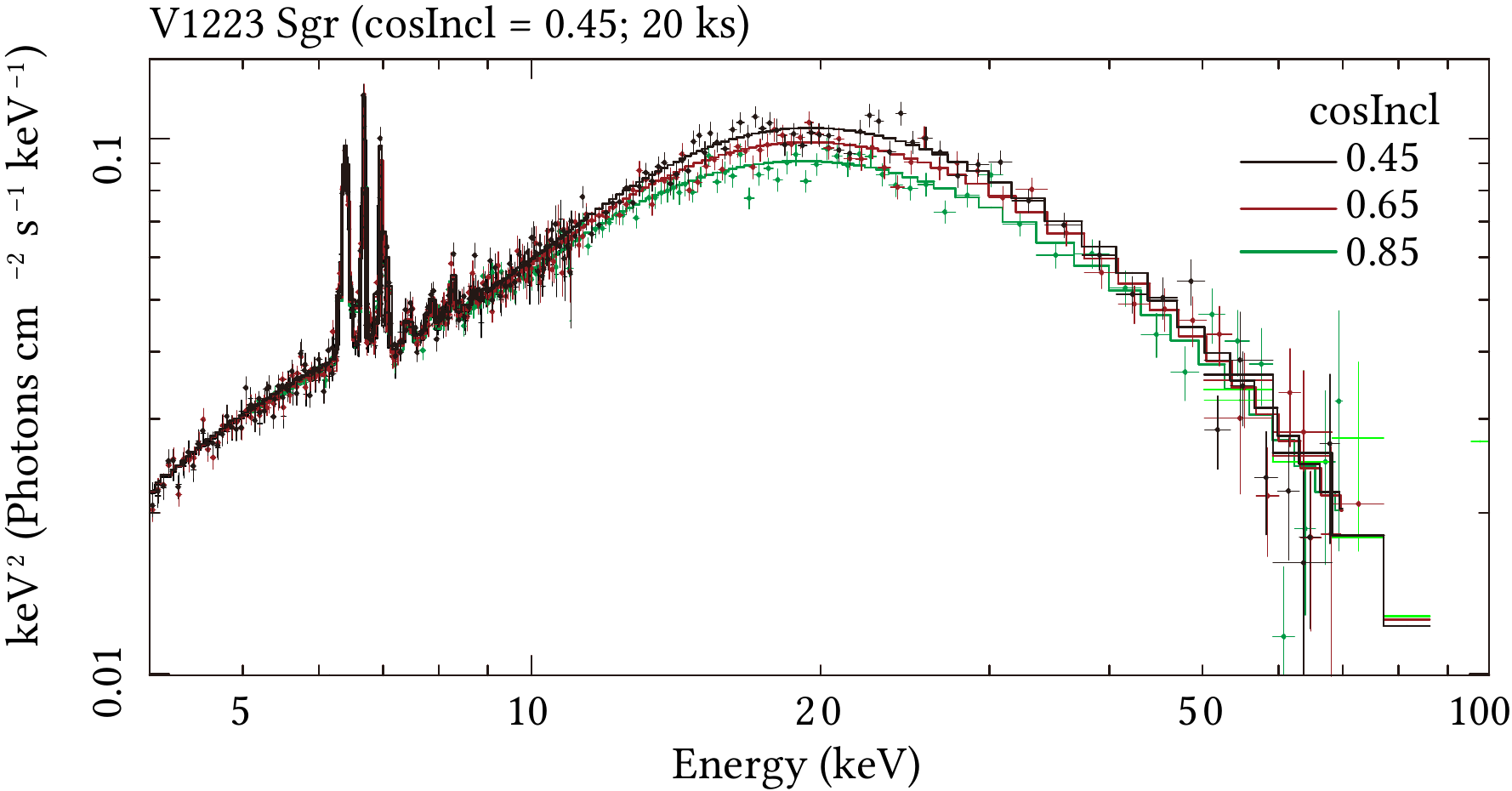}
\caption{Faked SXI+HXI $\nu F\nu$ spectra of V1223 Sgr. A total exposure of 100~ks is assumed. The presented spin-phase-resolved spectra are calculated with inclination angles (against the accretion column) $\cos\theta=0.45$, 0.65, and 0.85 each with 20~ks (i.e. 100~ks exposure is divided into 5 spin phases).}
\label{figure:fake_cosIncl_0.45_0.65_0.85}
\end{center}
\end{figure}

Figure \ref{figure:fake_cosIncl_0.45_0.65_0.85} shows spin-resolved
simulation spectra of V1223 Sgr. 
Spectral model of \citet{yuasa2013} has been convolved with 
the reflection model \verb|reflect| assuming a large reflector that covers
50\% of solid angle viewed from the PSR. In this simulation, each spin-phase
has net exposure of 20 ks (i.e. 100-ks observation is tentatively divided into five spin phases).
Count rates in the HXI energy band are tabulated in Table \ref{table:v1223_hxi_countrate}.

Although differences between individual spin phases are not necessary
obvious, the count rates can be used as a nice estimator of reflection fraction, and
variation of them may constrain angular coverage of reflecting material (WD surface).

\begin{table}[htb]
\begin{center}
\caption{Simulated count rates of V1223 Sgr in the hard X-ray band.}
\label{table:v1223_hxi_countrate}
\small{
{\renewcommand\arraystretch{1.25}
\begin{threeparttable}
\begin{tabular}{ccccc}
\hline
$\cos\theta^1$ & 0.25 & 0.45 & 0.65 & 0.85 \\
Count rate$^2$ & $1.334\pm0.008$ & $1.456\pm0.009$ & $1.550\pm0.009$ & $1.605\pm0.009$\\
\hline
\end{tabular}
\begin{tablenotes}\footnotesize
\item[1] Inclination between the line of sight and the PSR vertical direction (i.e. normal vector of the WD surface where accreting gas lands).
\item[2] Total $10-70$~keV count rate expected in the two HXI.
\end{tablenotes}
\end{threeparttable}

}
}
\end{center}
\end{table}

\subsection{Beyond Feasibility}
Based on previous studies, V1223 Sgr has a WD mass of $\sim0.7-0.8~M_{\odot}$.
Energies of fluorescent Fe K$\alpha$ line emitted at the WD surface should be affected by
the gravitational redshift of an order or 1~eV. Although this gravitational redshift should be 
cross-checked by observing CVs with heavier WDs to overcome systematic uncertainty of SXS energy scale (see below), if we can measure redshift amount in V1223 Sgr, we will be able to improve reliability of our WD gravitational potential estimation, or mass estimation, supported by this which is independent from other measurable quantities such as shock temperature or free-fall velocity of pre-shock gas.

\section{X-ray emission region in non-magnetic CVs}

%

\subsection{Background and Previous Studies}

Observers have long assumed that the Keplerian accretion disk in 
non-magnetic CVs extend down to the white dwarf surface, and the
boundary layer (BL) between the Keplerian disk and the white dwarf is
the likely site of much of the X-rays we observe \citep{PR1985}.
The detailed structure of the boundary layer, unfortunately, remains
poorly understood.  
In high accretion rate cases, the
boundary layer is expected to become optically thick, which should
make it a soft X-ray source (say 20 eV blackbody), not a hard
X-ray source.  Yet, high accretion rate non-magnetic CVs are observed
to emit hard X-rays.  The origin of these hard X-rays, in systems
that should have an optically thick boundary layer, is a major
unanswered question. As proposed in \citet{ishidaetal2009}, an accretion
disk wind, which is a common feature of high accretion rate disks in CVs,
may be connected with the hard X-ray emission in high accretion rates.
At present, however, the study of wind in X-ray data is at a relatively early stage.
Moreover, following the standard accretion disk \cite{SS1973},
half of the gravitational energy is released in the accretion disk and, hence, the other half is released in BL.
Observations in the extreme-ultraviolet band of SS Cyg and VW Hyi, 
however, revealed that the fractional energy radiated 
from BL is only $<$ 10\% of the disk luminosity (\citealt{Mauche1991}, \citeyear{Mauche1995}). 
According to classical theory, the temperature of BL in outburst is predicted to be 
2--5 $\times 10^5$~K \citep{Pandel2005},
whereas the temperature estimated by ultraviolet and optical emission lines 
is constrained to a significantly lower range of 5--10 $\times 10^4$~ K \citep{HoareDrew1991}.
These discrepancies may be resolved if we assume that BL is terminated not on the static white dwarf surface, 
but on a rapidly rotating accretion belt on the equatorial surface of the white dwarf 
(\citealt{Paczynski1978}; \citealt{KippenhahnThomas1978}).
Suggestions about the accretion belt, rotating at a speed close to the local Keplerian velocity, 
have been reported from a few non-magnetic CVs in outburst 
(\citealt{Huang1996}; \citealt{Sion1996}; \citealt{Cheng1997}; \citealt{Szkody1998}).

In addition, there is a question of whether the disk reaches the
surface in the case of low accretion rate disks.  Dwarf novae are
large subclass of non-magnetic CVs in which the mass transfer rate
from the secondary is low, the disk is cold and dim most of the
time (quiescence) with occasional outbursts when the disk is hot
and bright.  This is generally interpreted in the framework of
the disk instability model, or DIM \citep{lasota2001}. However, an essential
feature of DIM is that the accretion rate through a disk is highly
dependent on the radial distance from the white dwarf: somewhat
high near the outer edge, very low at the inner edge, hence matter
accumulates in the disk throughout the quiescent interval.  The basic
version of the DIM therefore predicts an extremely low accretion rate
onto the white dwarf during quiescence.  The X-ray luminosity
of quiescent dwarf novae immediately disproves this.  One possible
modification of DIM is that a central hole develops in quiescent
dwarf nova.

\subsection{Prospects \& Strategy}
\begin{figure}[htb]
\begin{center}
\includegraphics[width=\hsize, bb= 9 9 471 178]{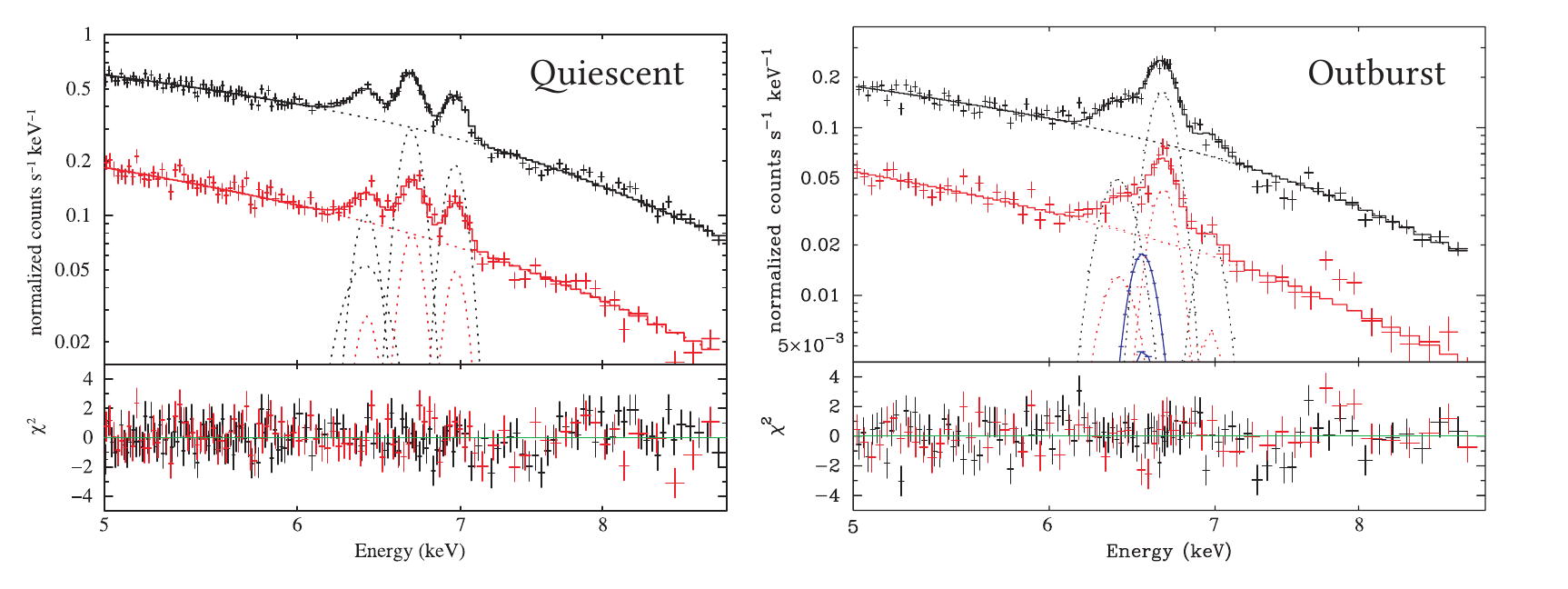}
\caption{{\it Suzaku} XIS FI (black) and BI (red) spectra of SS Cyg in quiescence (left; 39~ks) and outburst (right; 56~ks). Note that the Fe K$\alpha$ line structures are apparently different in the two states. Adopted from \citet{ishidaetal2009}. }
\label{figure:sscyg_xis}
\end{center}
\end{figure}

A comparison of X-ray spectra of a dwarf nova in quiescence\footnote{Quiescent and outburst in this section refer to those in optical wavelength. In conventional X-ray wavelength, e.g. in $2-10$~keV, photon flux in quiescent state is higher than that of outburst state.} and in outburst is highly instructive, and has been done with many X-ray satellites,
most recently with {\it Suzaku} \citep{ishidaetal2009}. Figure \ref{figure:sscyg_xis}
presents the XIS spectra taken during the quiescence and outburst states which
have apparently distinctive Fe K$\alpha$ line profiles.

During quiescence, the fluorescent Fe K$\alpha$ line is composed of two
components. The narrow component is interpreted as due to reflection off
the WD surface and the broad component due to reflection off the inner disk.
Given the equivalent widths of these component, the primary emission is
from the BL and there is no room for a central hole in the disk.  This
important conclusion can be confirmed, and the quantitative results
refined, with a relatively short {\it ASTRO-H}
observation utilizing the high spectral resolution SXS.

\begin{figure}[htb]
\begin{center}
\includegraphics[width=0.7\hsize, bb= 0 0 354 84]{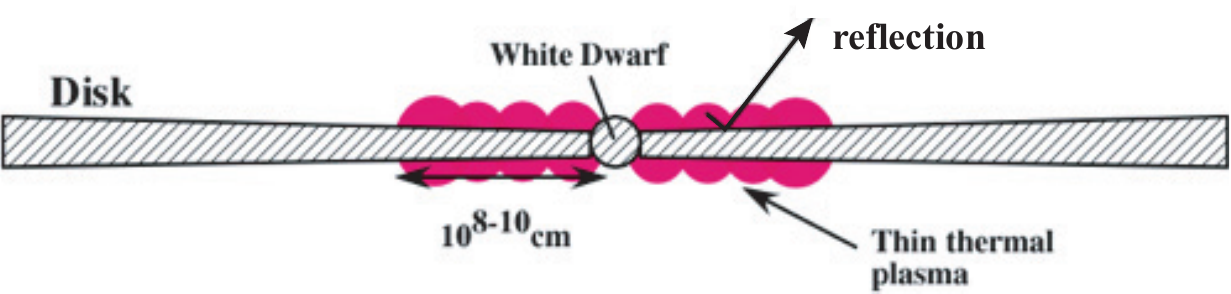}
\caption{A schematic view of accretion disk and corona of SS Cyg in the outburst state. The pink bubble-like object labeled with "Thin thermal plasma" is thought to be the origin of hard X-ray continuum and the broadened Fe K$\alpha$ lines. Adopted from \citet{ishidaetal2009}.}
\label{figure:sscyg_illustration}
\end{center}
\end{figure}

\begin{figure}[htb]
\begin{center}
\includegraphics[width=0.6\hsize, bb= 0 0 320 173]{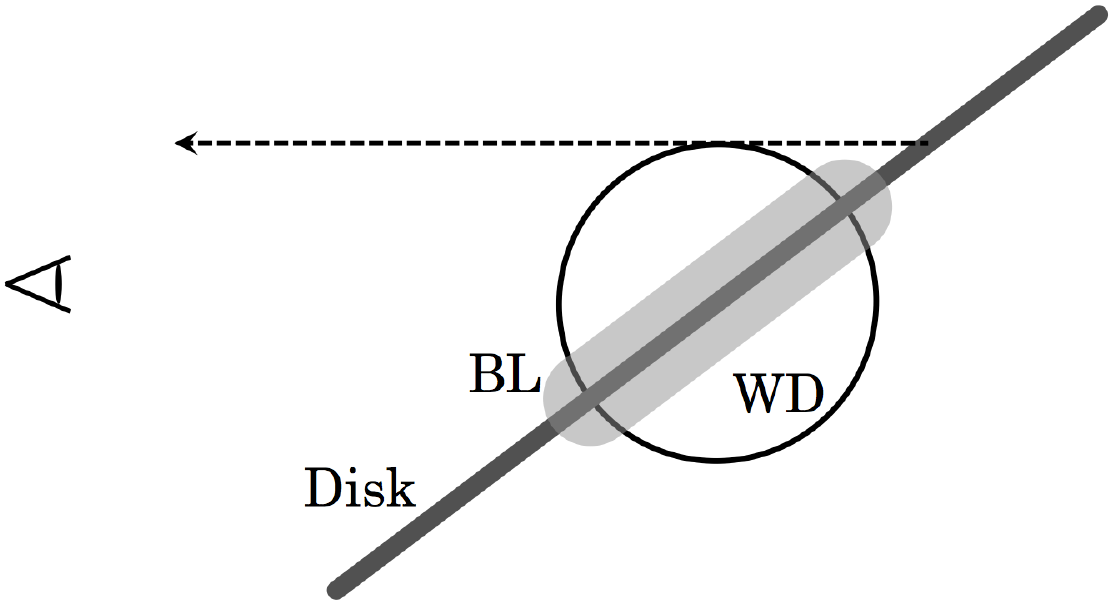}
\caption{A schematic view of occultation of the accretion disk or BL by the WD.}
\label{figure:non_magCV}
\end{center}
\end{figure}

In contrast, the fluorescent line is much broader during outburst.
\citet{ishidaetal2009} inferred from this that
the Fe K$\alpha$ lines and hard X-ray continuum are emitted from optically thin
thermal plasma somewhere above the accretion disk (namely disk corona) as
illustrated in Figure \ref{figure:sscyg_illustration}, and the Fe K$\alpha$ line profiles
are double-peaked due to Doppler red and blue shifts caused by rapid rotation of
the accretion disk and the corona. 
Moreover, if the system has moderate inclination and the plasma places close enough to the WD,
a part of X-ray is occulted by the WD as figure \ref{figure:non_magCV}
and the double-peaked lines are additionally distorted.

By precisely measuring the line profiles with the SXS, we will be able to determine
the line-of-sight velocity of the disk corona and geometrical size of the corona (from
equivalent width of fluorescence Fe K$\alpha$). The velocity can be related with
Keplerian rotation speed, and thus, the location of the corona will be estimated
enabling detailed study of disk corona which is not well understood in CVs.

\subsection{Targets \& Feasibility}
We believe that SS Cyg, which is the most widely studied dwarf novae, is the best target when studying disk corona and disk reflection with the SXS. Physical parameters of the system have been precisely measured as tabulated in Table \ref{table:sscyg_parameters}, and therefore, we will be able to precisely correlate Doppler speed measured with the SXS to Keplerian motion estimating the innermost  corona radius (and probably innermost disk radius). An outburst period of this system is about $50$~days lasting $\sim10$~days which is suitable for performing a time-constrained observation with {\it ASTRO-H}.

\begin{table}[htb]
\begin{center}
\caption{System parameters of SS Cyg (\citealt{ishidaetal2009} and references therein,
with an updated distance from \citealt{MJetal2013}).}
\label{table:sscyg_parameters}
\small{
{\renewcommand\arraystretch{1.25}
\begin{tabular}{ccccc}
\hline
Parameter & Value & \ & Parameter & Value\\
\hline
$M_{\mathrm{WD}}$ & $1.19\pm0.02~M_{\odot}$ & & $M_{\mathrm{secondary}}$ & $0.704\pm0.002~M_{\odot}$\\
$P_{\mathrm{orbital}}$ & 6.6~hr & & $P_{\mathrm{outburst}}$ & $\sim50$~days\\
Inclination & $37^\circ\pm5^\circ$ & & Distance & $114\pm2$~pc\\
\hline
\end{tabular}
}
}
\end{center}
\end{table}

\begin{figure}[!h]
\begin{center}
\includegraphics[width=0.6\hsize, bb=0 0 670 723]{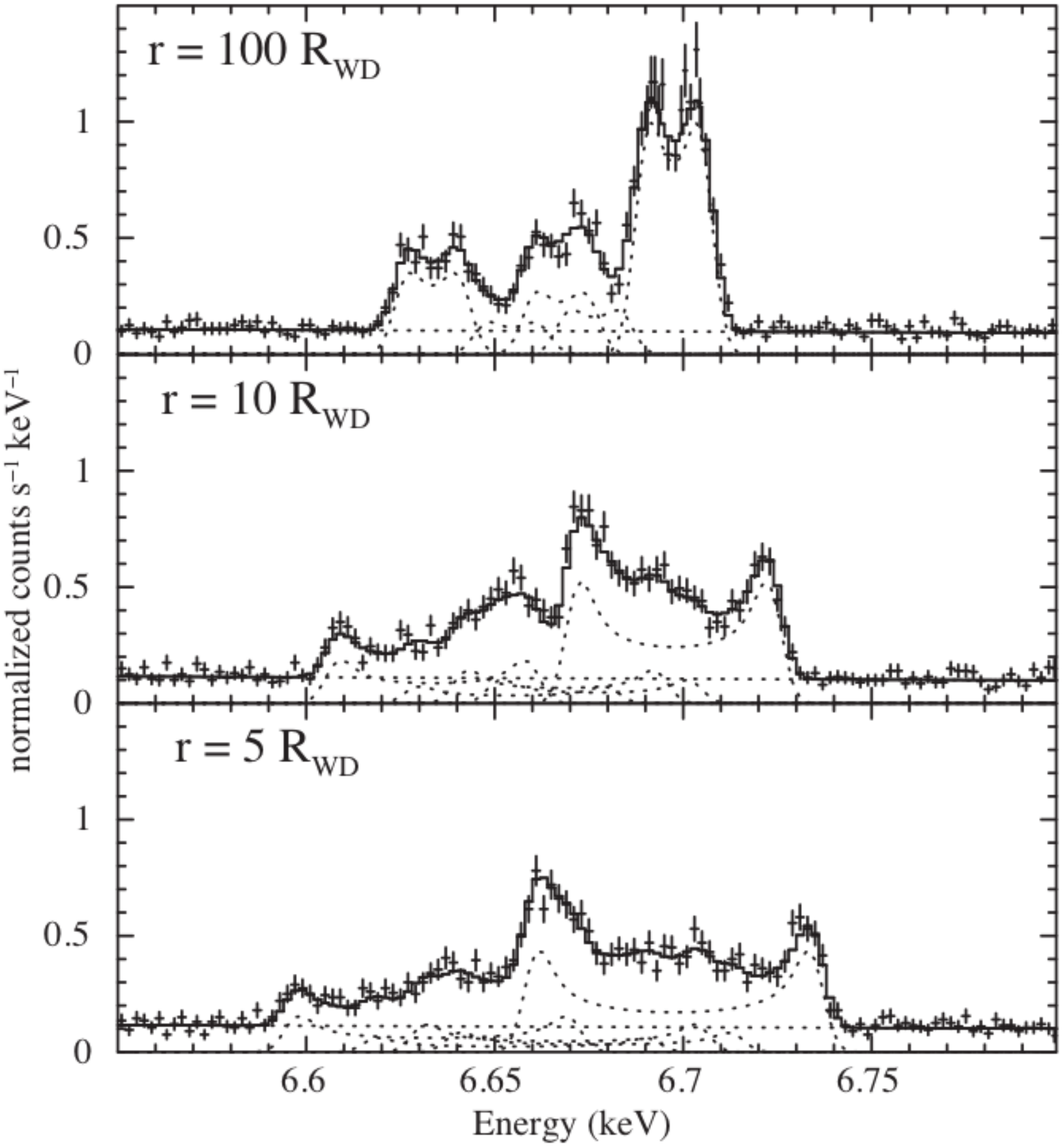}
\caption{Simulated SXS spectra of SS Cyg. Assumed exposure is 100-ks each, and three spectra were calculated for disk inner radius $r$ of $100~R_{\mathrm{WD}}$, $10~R_{\mathrm{WD}}$, and $10~R_{\mathrm{WD}}$ from the top to the bottom.}
\label{figure:sscyg_simulation}
\end{center}
\end{figure}

\begin{figure}[!h]
\begin{center}
\includegraphics[width=0.6\hsize]{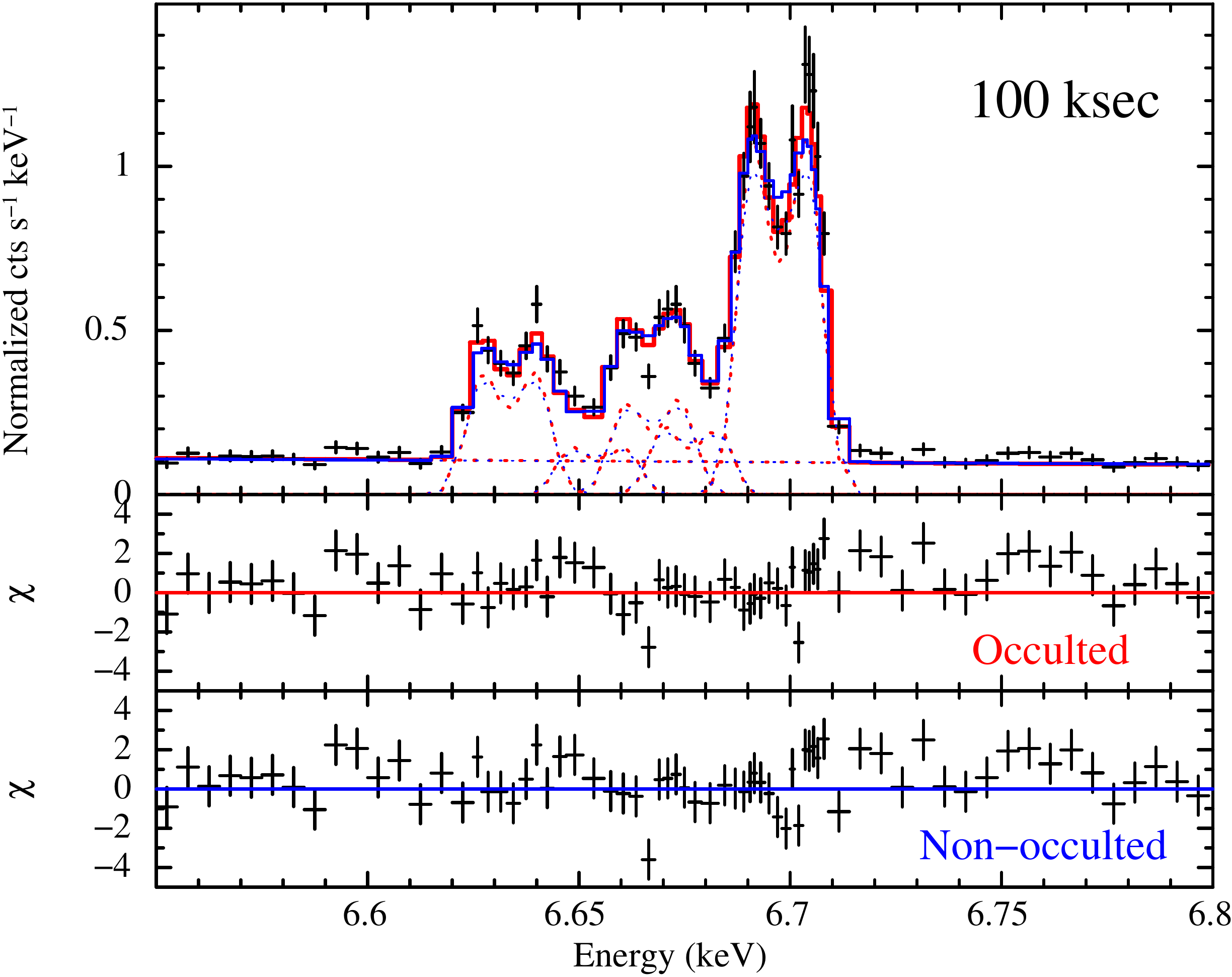}
\caption{Top panel shows a simulated spectrum (black) for 100 ks exposure 
fitted with the occulted (red) and non-occulted (blue) 1/10 Keplerian models.
The middle and the bottom panel show residuals for the two models.}
\label{figure:sscyg_simulation_eclipse}
\end{center}
\end{figure}

Figure \ref{figure:sscyg_simulation} shows simulated Fe K$\alpha$ spectra of SS Cyg in outburst assuming 100 ks exposure and spectral model proposed in \citet{ishidaetal2009}. Three separate innermost corona radii were assumed thus resulting different Doppler broadening of the lines. By fitting the simulated spectra with the model, we will be able to determine the disk inner radius within $\pm5\%$ as well as plasma temperature of the corona based on the He-like and H-like line ration. In this simulation, we used the SXS instrumental response files as follow:
\begin{itemize}
\item \verb|ah_sxs_7ev_basefilt_20090216.rmf| (energy redistribution file),
\item \verb|sxt-s_120210_ts02um_intallpxl.arf| (auxiliary response file), and 
\item \verb|sxs_cxb+nxb_7ev_20110211_1Gs.pha| (background spectral file).
\end{itemize}
We also simulated spectra emitted by the plasma rotating with 1/10 of  Kelper velocity at WD surface 
which is occulted by the WD with 100 ks exposure. 
The plasma distances from the WD were assumed to be the 0.1 times of the white dwarf radius.
Figure \ref{figure:sscyg_simulation_eclipse} shows a comparison of fitting results of
the simulated spectra with the occulted and non-occulted 1/10 Keplerian models. 
The larger residual is left with the non-occulted 1/10 Keplerian model. 
The no more than 7\% accuracy for the distance between the plasma and the WD surface
was obtained by fitting with the occulted 1/10 Keplerian model thawing the disk velocity.
The significance of consideration of the occultation is $<$ 99.99\%. 
Note that the uncertainties of instrument responses are not important 
for the spectral fitting to the distorted lines because the concerning energy ranges are narrow.

Thus, an SS Cyg observation will fully exploit the SXS capabilities, and constrain physical parameters of the disk corona which is not well studied in previous and on-going X-ray missions. Since hot disk corona is thought to be a common feature of accreting stellar
black holes, development of understanding of disk corona in accreting non-magnetic
CVs will infer ubiquity of disk corona in accretion physics independent of mass of accreting
objects.

\section{Anisotropic Radiative Transfer of Resonance photons}
\subsection{Background and Previous Studies}

Resonance scattering by heavy ions can play important
rules in radiative transfer, because the cross section
of the resonance scattering is two orders of magnitude larger than
that of Compton scattering in the X-ray band. In other words,
resonance photons carry additional information regarding the physical  conditions of the plasma.
For example, optically thin hot plasmas in clusters of galaxies are
optically thick for resonance scattering, but the resonance-scattering
process is suppressed if a significant turbulent motion exists 
in the plasmas. Therefore, the equivalent widths of resonance lines can
be used to quantify the amount of turbulence.
The post-shock region (PSR) of magnetic CVs is another example in which
resonant scattering can play a significant role.
Typically, an optical depth of an accretion column in a magnetic CV for 
Compton scattering is of order $\tau_{\it CMP} \sim 10^{-1}$.
However, the optical depth for resonance scattering process
is of order $\tau_{\rm RS} \sim 10^2$.
Therefore, resonance photons carry (1) the information about the geometry
of the column and/or (2) internal information of the plasma (velocity, 
temperature, and density gradients), through radiative transfer
in the accretion column.
Compared with plasmas in clusters of galaxies, one advantage of magnetic CV 
observations is that we can scan various viewing angles to the plasma
using the spin of the WD.

\begin{figure}[hbt]
\begin{center}
\includegraphics[width=0.5 \textwidth]{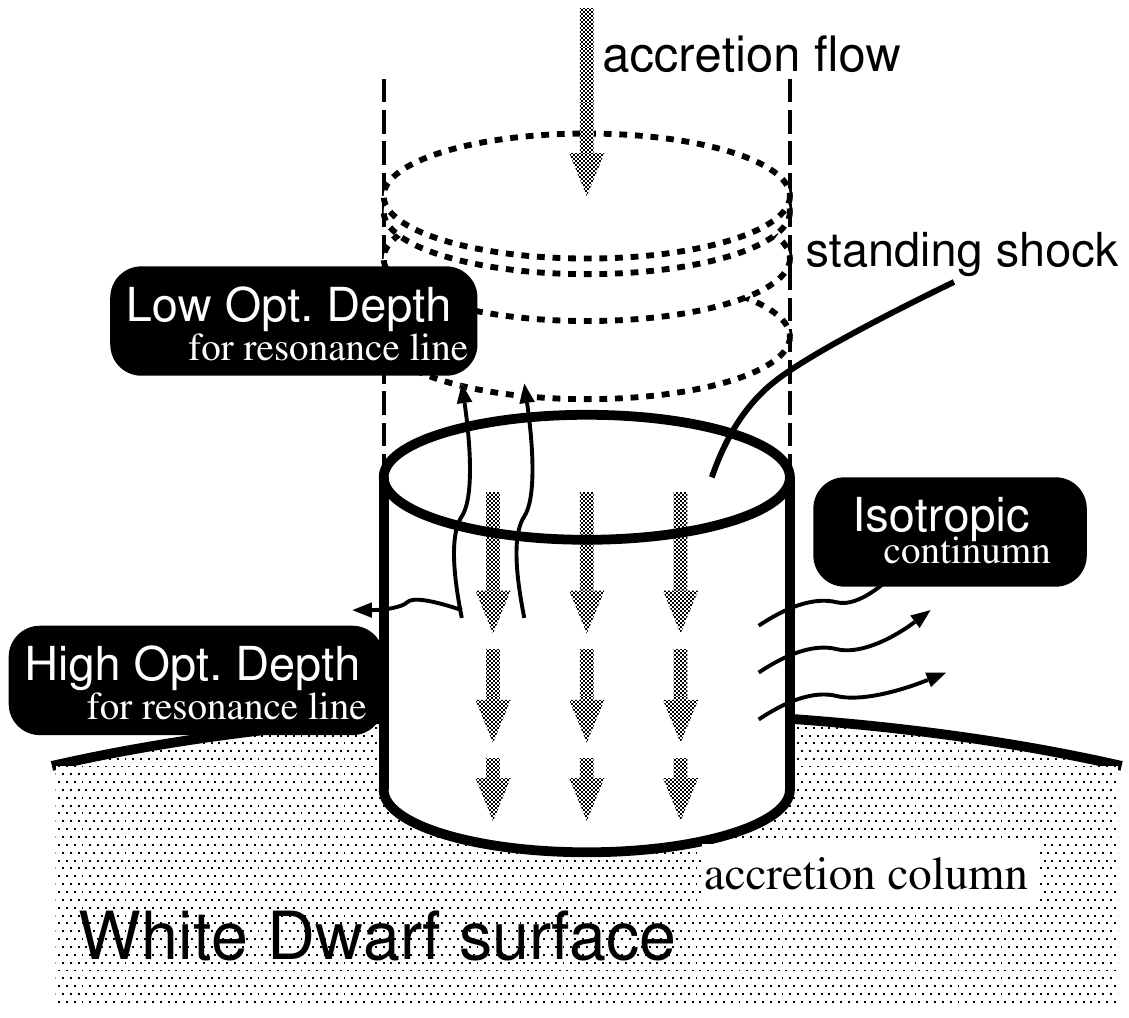}
\caption{Schematic view of the accretion column of magnetic CVs and their optical
depths to Compton and resonance scatterings. Adopted from \citet{teradaetal2001}.}
\label{figure:resonance_schematic}
\end{center}
\end{figure}

The idea of anisotropic radiative transfer of resonance photons 
in the accretion columns of polar type magnetic CVs was originally proposed briefly in 
\citet{teradaetal1999}.
Only the surface of the accretion column is observed by resonance 
lines due to the high opacity, whereas thermal photons show isotropic 
emission (i.e., whole volume is seen). 
Therefore, the equivalent width of resonance line will be enhanced 
when we observe the accretion column from the pole-on direction 
when it has flat coin-like shape.
In addition to such geometrical effects, another collimation effect 
due to velocity gradient in the column is also expected \citet{teradaetal2001}.
As shown in the schematic view of the accretion column 
in Figure \ref{figure:resonance_schematic}, 
the velocity of the bulk flow of the gas has a vertical gradient, 
and thus the resonance scattering optical depth $\tau_{\rm RS}$ 
is reduced in the vertical direction, while it remains optically thick
in the horizontal direction.
Therefore, the resonance photons preferentially escape in 
the vertical direction of the column.
Note that the anisotropy in $\sigma_{\rm RS}$ can be suppressed 
by the thermal broadening effect on $\sigma_{\rm RS}$; 
the vertical structure of the temperature should be considered 
in the radiative transfer.
These effects are numerically clarified using Monte Carlo 
simulations of the radiative transfer in the accretion column \citep{teradaetal2001}.

Observationally, anisotropic transfer of resonance photons was 
marginally confirmed by phase-resolved analyses of Fe K lines with 
{\it ASCA} observation of the polar, V834 Cen \citep{teradaetal2001}
and tested for other 18 magnetic CVs observed with {\it ASCA} \citep{teradaetal2004}.
These observations indicate that this effect is significant in polars,
but not in intermediate polars.  The likely explanation for the difference
between the two subclasses of magnetic CVs is the geometry. Intermediate
polars, accretion is via a partial disk, in the form of ``accretion curtains''
that cover large ranges in magnetic latitude, and photons can easily escape
preferentially perpendicular to the curtains.

However, it is important to note that the relatively poor energy resolution of the {\it ASCA} SIS
limited our ability to separate pure resonance lines from the blend of Fe K lines; 
the Fe XXVI K$_\alpha$ line is the pure resonance line but is week and 
is difficult to be separated from Fe XXV K$_\alpha$ and/or fluorescent Fe lines.
Needless to say, it is impossible to distinguish resonance (6.698 keV), 
inter-combination (6.673 keV), and forbidden (6.637 keV) lines 
of Fe XXV K$_\alpha$ blends with CCD.
As for lighter elements (O, Mg, Si, S, etc), the anisotropy effect 
by the bulk velocity gradient is expected to be strongly suppressed 
by the thermal Doppler broadening, 
and thus the effect could not be tested by grating observations 
with {\it Chandra} or {\it XMM-Newton}.
Therefore, only {\it ASTRO-H} SXS observations can test
the anisotropic radiative transfer of resonance photons.

\subsection {Prospects \& Strategy}
\begin{figure}[ht]
\begin{center}
\includegraphics[width=0.5 \textwidth]{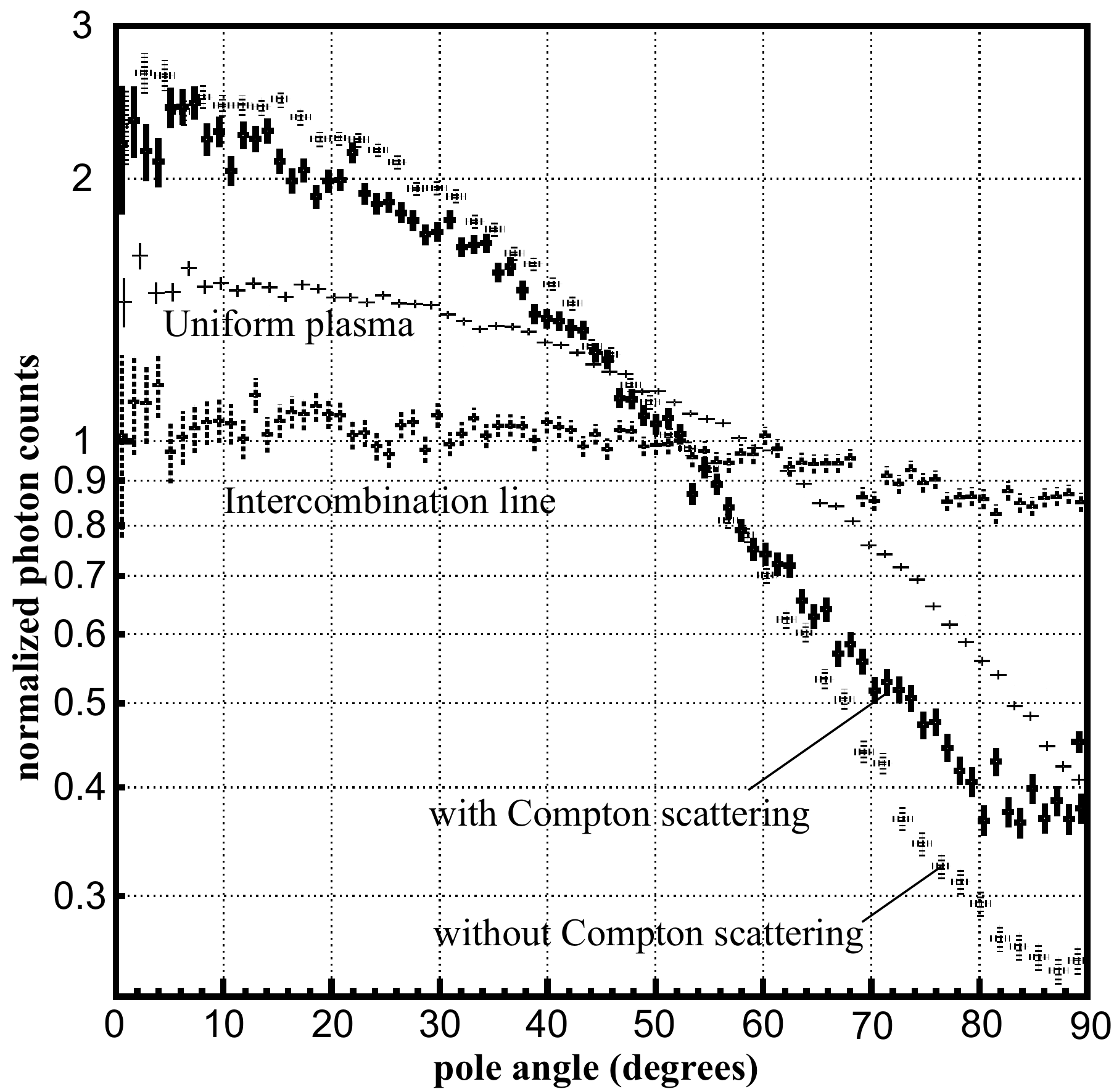}
\caption{Angular distributions of resonance and inter-combination lines
of Fe XXV K$_\alpha$. Adopted from \citet{teradaetal2001}.}
\label{figure:resonance_sim_dist}
\end{center}
\end{figure}

According to the Monte Carlo simulation of radiative transfers of 
resonance and continuum photons in the accretion column by 
\citet{teradaetal2001}, we can expect a factor of 2 or 3 enhancement at maximum.
Figure \ref{figure:resonance_sim_dist} shows a distribution of photons 
as a function of the angle from the vertical axis of the column 
(pole angle; 0 and 90 degrees mean pole-on and side-on views).
For inter-combination lines, which must show isotropic emission, 
the photon flux has almost no dependency on the pole angle, as expected.
On the other hand, resonance photons is enhanced to the pole-on direction 
by factor 2 or 3, as illustrated in the vertical axis of the figure. 
For reference, the figure shows plots without Compton scattering or 
without vertical gradient of the bulk velocity.

By precise measurements of enhancement of resonance photons with the SXS, 
we can test the assumption in the simulation on the structure of the 
plasma (i.e., the vertical gradient of the bulk velocity, temperature, 
and densities) independently from the measurements of the continuum 
or other lines described in the previous sections.

\subsection{Targets \& Feasibility}
\begin{figure}[ht]
\begin{center}
\includegraphics[width=0.8 \textwidth]{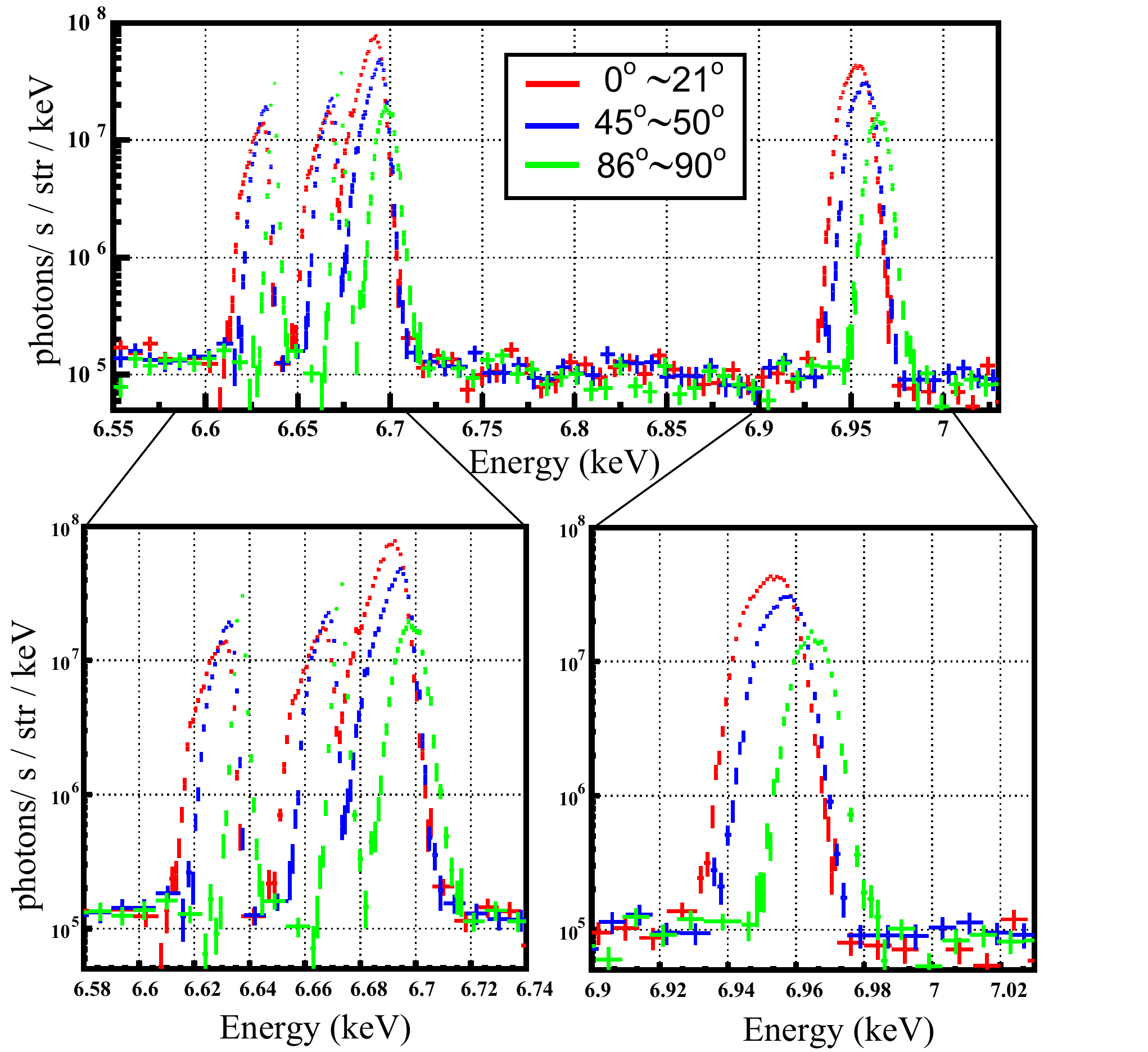}
\caption{Phase resolved energy spectra for iron K$_\alpha$ lines calculated
by the Monte Carlo simulation \citep{teradaetal2001}. Adopted from \citet{teradaphdthesis}.}
\label{figure:resonance_sim_spec}
\end{center}
\end{figure}

Figure \ref{figure:resonance_sim_spec} shows simulated phase-resolved 
X-ray spectra around Fe K band by the Monte Carlo simulation 
\citep{teradaetal2001}, under the assumption that we can observe the pole 
angle from 0 to 90 degrees.
Thus, we can expect the enhancement on the resonance line, 
as a function of Doppler shift by the vertical bulk motion of the gas. 
This is the direct verification of the anisotropic resonance-scattering 
effect.

We must select a target that allows us to observe a wide range of pole
angles, using the tabulated values of magnetic colatitude $\beta$ and
the inclination angle $i$.
They are summarized in Table 4 of \citet{teradaetal2001};
by this criterion, the best candidates are VY For and EK UMa;
AM Her, V834 Cen, and GG Leo (=RX~J1015+09), are also good candidates.
Once we also considering the known hard X-ray fluxes, the latter three are
more suitable targets, however.
In addition, polars are known to have low states in which the X-ray
flux can drop by an order of magnitude or more. An observation during
a low state will likely be photon-starved and will not allow us to
take advantage of the spectral resolution of the {\it ASTRO-H} SXS.

\clearpage
\begin{multicols}{2}
\end{multicols}
\end{document}